\documentclass[superscriptaddress,amsmath,onecolumn,amssymb]{revtex4}
\usepackage{xr}

\usepackage{subfigure}
\usepackage{amsmath}
\usepackage{pgfplots}
\pgfplotsset{compat=1.5}

\usepackage{color}

\usepackage{xcolor}
\usepackage{makecell}
\usepackage{multirow}
\usepackage{placeins}
\usepackage{soul}
\usepackage{color,xcolor}
\usepackage[colorlinks,linkcolor=red,anchorcolor=blue,citecolor=blue,urlcolor=blue]{hyperref}
\usepackage{cleveref}

\begin{document}

\title{General Quantum Bernoulli Factory: Framework Analysis and Experiments}

\author{Yong Liu}\thanks{These authors contribute equally to this work.}
\affiliation{Institute for Quantum Information \& State Key Laboratory of High Performance Computing, College of Computer science and technology, National University of Defense Technology, Changsha 410073, China}
\affiliation{College of Information and Communication, National University of Defense Technology, Xi'an, 710006, China}

\author{Jiaqing Jiang}\thanks{These authors contribute equally to this work.}
\affiliation{Institute of Computing Technology, Chinese Academy of Sciences, Beijing 100190, China}
\affiliation{University of Chinese Academy of Sciences, Beijing 100049, China}

\author{Pingyu Zhu}
\affiliation{Institute for Quantum Information \& State Key Laboratory of High Performance Computing, College of Computer science and technology, National University of Defense Technology, Changsha 410073, China}
\author{Dongyang Wang}
\affiliation{Institute for Quantum Information \& State Key Laboratory of High Performance Computing, College of Computer science and technology, National University of Defense Technology, Changsha 410073, China}
\author{Jiangfang Ding}
\affiliation{Institute for Quantum Information \& State Key Laboratory of High Performance Computing, College of Computer science and technology, National University of Defense Technology, Changsha 410073, China}
\author{Xiaogang Qiang}
\affiliation{Institute for Quantum Information \& State Key Laboratory of High Performance Computing, College of Computer science and technology, National University of Defense Technology, Changsha 410073, China}
\affiliation{National Innovation Institute of Defense Technology, AMS, 100071 Beijing, China}
\author{Anqi Huang}
\affiliation{Institute for Quantum Information \& State Key Laboratory of High Performance Computing, College of Computer science and technology, National University of Defense Technology, Changsha 410073, China}
\author{Ping Xu}
\affiliation{Institute for Quantum Information \& State Key Laboratory of High Performance Computing, College of Computer science and technology, National University of Defense Technology, Changsha 410073, China}

\author{Jialin Zhang}
\affiliation{Institute of Computing Technology, Chinese Academy of Sciences, Beijing 100190, China}
\affiliation{University of Chinese Academy of Sciences, Beijing 100049, China}
\author{Guojing Tian}
\affiliation{Institute of Computing Technology, Chinese Academy of Sciences, Beijing 100190, China}
\affiliation{University of Chinese Academy of Sciences, Beijing 100049, China}

\author{Xiang Fu}
\affiliation{Institute for Quantum Information \& State Key Laboratory of High Performance Computing, College of Computer science and technology, National University of Defense Technology, Changsha 410073, China}
\author{Mingtang Deng}
\affiliation{Institute for Quantum Information \& State Key Laboratory of High Performance Computing, College of Computer science and technology, National University of Defense Technology, Changsha 410073, China}
\author{Chunqing Wu}
\affiliation{Institute for Quantum Information \& State Key Laboratory of High Performance Computing, College of Computer science and technology, National University of Defense Technology, Changsha 410073, China}

\author{Xiaoming Sun}
\email{sunxiaoming@ict.ac.cn}
\affiliation{Institute of Computing Technology, Chinese Academy of Sciences, Beijing 100190, China}
\affiliation{CAS Center of Excellence in Topological Quantum Computation, University of Chinese Academy of Sciences, Beijing 100190, China}

\author{Xuejun Yang}
\affiliation{Institute for Quantum Information \& State Key Laboratory of High Performance Computing, College of Computer science and technology, National University of Defense Technology, Changsha 410073, China}
\author{Junjie Wu}
\email{junjiewu@nudt.edu.cn}
\affiliation{Institute for Quantum Information \& State Key Laboratory of High Performance Computing, College of Computer science and technology, National University of Defense Technology, Changsha 410073, China}

\begin{abstract}
  The unremitting pursuit for quantum advantages gives rise to the discovery of a quantum-enhanced randomness processing named quantum Bernoulli factory (QBF). This quantum enhanced process can show its priority over the corresponding classical process through readily available experimental resources, thus in the near term it may be capable of accelerating the applications of classical Bernoulli factories, such as the widely used sampling algorithms. In this work, we provide the framework analysis of the QBF. We thoroughly analyze the quantum state evolution in this process, discovering the field structure of the constructible quantum states. Our framework analysis shows that naturally, the previous works can be described as specific instances of this framework. Then, as a proof of principle, we experimentally demonstrate this framework via an entangled two-photon source along with a reconfigurable photonic logic, and show the advantages of the QBF over the classical model through a classically infeasible instance. These results may stimulate the discovery of advantages of the quantum randomness processing in a wider range of tasks, as well as its potential applications.

\end{abstract}

\keywords{quantum Bernoulli factory, quantum randomness processing, photonic logic}

\date{\today}

\maketitle

\section{Introduction}

Quantum computers are trusted to be advanced over the classical machines in information processing, because of the counterintuitive features of quantum mechanics~\cite{Nielsen2002,Bennett2014,Gisin2007,Ekert1991,Grover1996,Shor1997}. Conventional efforts for quantum computing achieve the milestone named ``quantum computational advantages'' by building large-scale controllable computing devices~\cite{Wu2018,GouLiu2019,Preskill2012,Arute2019,Zhong2020}. However, there exist tasks where the advantages of quantum system over the classical system can be realized without building large quantum systems, such as the quantum Bernoulli factory~\cite{Dale2015}.

Consider the following problem: given a biased coin with unknown probability $p$ for head, by tossing this coin, can we exactly simulate a coin with probability $f(p)$ to get a head? For example, if $f(p)=p^2$, we can toss the coin twice and claim a head if both the results are head. In this case, we obtain a classical Bernoulli factory (CBF) for function $f(p)=p^2$~\cite{Nacu2005,Latuszynski2009,Thomas2011}. Any function $f(p)$ is ``constructible'' if there exists such a process that its success probability is $f(p)$ regardless of the value of $p$, which also corresponds to simulate a biased coin with probability $f(p)$ for head (denoted by $f(p)$-coin for simplicity). Importantly, the process should be expected to finish in finite steps, which means the expected number of coins consumed in the process is also finite. This type of processes is flexible and efficient in generating a variety of binary distributions, and have been applied in scenarios such as enhancing the Markov chain Monte Carlo process for intractable distributions~\cite{Krzysztof2011,Flegal2012,Goncalves2017_1,Vats2020}, exact simulations of diffusions~\cite{Goncalves2017_2,Blanchet2020} and the estimation for ocean circulations~\cite{Herbei2014}.

Unfortunately, for any classically constructible $f(p)$-coins, the function $f(p)$ is bounded by three conditions~\cite{Keane1994}: (i) The function should be continuous on its domain; (ii) The function should not reach 0 or 1 within its domain; (iii) The function should not approach 0 or 1 exponentially fast near any edge of its domain. Functions violating the conditions require infinite coin tossing if they are intended to be exactly constructed, such as the ``Bernoulli doubling'' function $f_\wedge(p)=2p, p\in[0,0.5]$, which is believed to be fundamental for other classical Bernoulli factory processes~\cite{Asmussen1992}.

To push this limit of the CBF, it can be generalized to a quantum version named quantum Bernoulli factory (QBF), which can accelerate the construction processes as well~\cite{Dale2015}. The QBF is a randomness processing task which applies quantum coins (or {\em quoins}) to produce classical coins. It starts from a group of identical quoins, each of which is actually a single-qubit state $|\psi_p\rangle=\sqrt{p}|0\rangle +\sqrt{1-p}|1\rangle$ where $p$ is the unknown parameter. With the quantum coherence and entanglement, the QBF is capable to produce strictly more results than those can be produced in CBF, including the Bernoulli doubling function, and the experimental demonstration of this phenomenon requires quite few physical resources~\cite{Yuan2016,Patel2019}.

However, the proposed works of the QBF mainly analyze the advantages that can be obtained by utilizing quantum coherence. Meanwhile, previous efforts only focused on specific cases such as the QBF for the Bernoulli doubling function, or analyzed the range of constructible quantum states in single-qubit cases, while the multi-qubit cases were not sufficiently studied~\cite{Jiang2018,Zhan2020}.

Addressing these problems, in this work, we completely analyze the evolution of multi-qubit quantum states in the QBF for classical coin generation, and found that the arbitrary constructible states can be characterized by a field structure. We then analyze the impacts of different capabilities of the quantum processors on the range of constructible coins and the construction performance. Finally, we experimentally realize the framework of QBF and show the advantages of quantum randomness process over its classical counterpart through a classically infeasible instance. Our work indicates the potentials of QBF in accelerating the applications where the CBF were already applied, and may motivate the interdisciplinary studies of quantum computing and randomness processing.

\section{Constructible states in the QBF}


The key to push the limits of the CBF is to construct proper quantum states. In QBF, the quantum states are evolved from several copies of the state $|\psi_p\rangle=\sqrt{p}|0\rangle+\sqrt{1-p}|1\rangle$, where $p$ is the unknown parameter corresponding to the $p$-coin in the CBF. The goal of this analysis is to find out whether a $p$-involved state (i.e. any amplitudes of these states are functions of $p$) can be obtained from $|\psi_p\rangle$, and a state is {\em constructible} if it can be transformed from $|\psi_p\rangle$ in finite steps with a nonzero probability. In each step, one can apply unitary operations or measurements. The states which already have been constructed can be used as auxiliary qubits.

Firstly, we briefly review the single-qubit cases~\cite{Jiang2018}. Generally, a single-qubit $p$-involved state can be represented by $|\psi_o\rangle=k_0(p)|0\rangle+k_1(p)|1\rangle$, where $k_0(p)$ and $k_1(p)$ are functions of $p$, satisfying $|k_0(p)|^2+|k_1(p)|^2=1$. We concentrate on the amplitude of the $|0\rangle$ basis, and rewrite the state as
\begin{equation}\label{eq:singlequbit}
|h(p)\rangle=c(p)\left(h(p)|0\rangle+|1\rangle\right),
\end{equation}
where $c(p)$ is the coefficient for normalization, and $h(p)=k_0(p)/k_1(p)$ is called the relative amplitude. 
It has been proved~\cite{Jiang2018} that a single-qubit state $|h(p)\rangle$ is constructible if and only if $h(p)$ belongs to the field $\mathbb{M}$ which is spanned by $p$-involved term $\sqrt{\frac{p}{1-p}}$ and the complex field, that is
\begin{equation}\label{eq:setM}
  \mathbb{M}=\left\{\frac{g_1(p)}{g_2(p)}\sqrt{\frac{p}{1-p}}+\frac{g_3(p)}{g_4(p)}\right\},
\end{equation}
where $g_i(p)$ are polynomials in $p$ with complex coefficients. Note that here our goal is to construct the state $|h(p)\rangle$ exactly with the unknown parameter $p$. Though we can approximate any well-behaved functions with polynomials, which however requires extremely large number of steps to reach a certain accuracy. Besides, since the relative amplitudes of the constructible quoins form a field, the unitary operations corresponding to the basic operations defined for this field provide a general approach to produce arbitrary constructible states. However, since the produced states are limited in single-qubit cases, it is unable to perform the joint measurement on the result state to obtain classical probability as done in ref~\cite{Dale2015, Patel2019}.

To tackle this issue, we generalize this result to multi-qubit cases. For a general QBF process, it produces an $n$-qubit state $|K(p)\rangle=\sum_{i=0}^{2^n-1}k_i(p)|i\rangle$ from several copies of $|\psi_p\rangle$ in finite steps with a nonzero probability, where $n\geq 1$ and $k_i(p)$ are functions of $p$, satisfying $\sum_{i=0}^{2^n-1}\left|k_i(p)\right|^2=1$. In each step, one can apply arbitrary unitary operations, or measurements on any part of this state.
Here we are using $|i\rangle$ to represent an $n$-qubit computational basis, with its binary form representing the states of all the qubits, i.e. $|i\rangle=|i_{n-1}i_{n-2}\cdots i_0\rangle$ where $i = \sum_{j=0}^{n-1}i_{j}\cdot 2^{j}$. 
We denote all the constructible states by the notion of {\em Bernoulli states}. We then rewrite the $n$-qubit Bernoulli state as
\begin{equation}\label{EQ:REWRITTEN}
|K(p)\rangle=c(p)\left(\sum_{i=0}^{2^n-2}h_i(p)|i\rangle+|2^n-1\rangle\right),
\end{equation}
where $h_i(p)=k_i(p)/k_{2^n-1}(p)$ is the relative amplitude, and $c(p)$ is used for normalization. 
Our result is the following theorem.

{\bf {\em Theorem.}} An $n$-qubit state $|K(p)\rangle=c(p)\left(\sum_{i=0}^{2^n-2}h_i(p)|i\rangle+|2^n-1\rangle\right)$ is constructible if and only if every $h_i(p), (i=0,1,\cdots,2^n-2)$ belongs to $\mathbb{M}$.

Here we show the key steps to prove this theorem, and the detailed proof can be found in the supplementary information. Let $\mathbb{S}$ be the set of the relative amplitudes of all constructible states, then our goal is to prove $\mathbb{S}=\mathbb{M}$. The necessity (that is $\mathbb{S}\subseteq\mathbb{M}$) is easy to show based on the conclusion of ref.~\cite{Jiang2018}. For the sufficiency (that is $\mathbb{M}\subseteq\mathbb{S}$), since any two amplitudes can be switched under specific unitary operations, in the following we focuses on $h_0(p)$ without lossing generality. Suppose that we have a multi-qubit state $|K(p)\rangle$ as described by eq.~(\ref{EQ:REWRITTEN}) and a single-qubit state $|l(p)\rangle=l(p)|0\rangle+|1\rangle$, then the target is to find a group of unitary operations to construct the following states:
\begin{equation}
\left\{
\begin{aligned}
  |K_1(p)\rangle&=c_1(p)\left(\frac{1}{h_0(p)}|0\rangle+\sum_{i=1}^{2^n-2}h_i(p)|i\rangle+|2^n-1\rangle\right)\\
  |K_2(p)\rangle&=c_2(p)\left(h_0(p)l(p)|0\rangle+\sum_{i=1}^{2^n-2}h_i(p)|i\rangle+|2^n-1\rangle\right)\\
  |K_3(p)\rangle&=c_3(p)\left((h_0(p)+l(p))|0\rangle+\sum_{i=1}^{2^n-2}h_i(p)|i\rangle+|2^n-1\rangle\right),\\
\end{aligned}
\right.
\end{equation}
where the $c_i(p)$ are the coefficients for normalization. These unitary operations only changes the relative amplitude of the $|0\rangle$ basis, and therefore respectively realizes the three field-operations including inversion, multiplication, and addition operation. This indicates that 
the set $\mathbb{S}$ is closed under these three operations, and therefore it is exactly the filed $\mathbb{M}$. 

\section{The framework of the QBF}

\begin{figure*}[b]
  \centering
  \includegraphics[width=\textwidth]{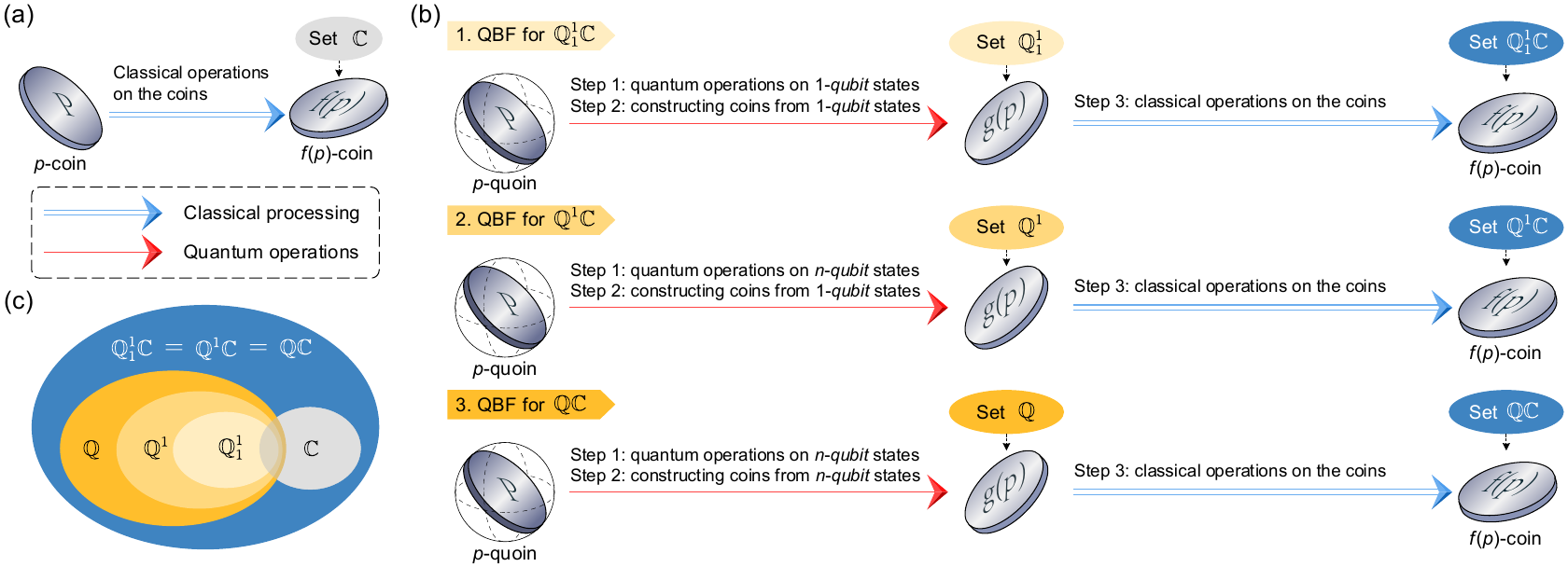}\\
  \caption{\footnotesize{{\bf The framework analysis of the QBF.} {\bf (a)} the classical Bernoulli factory (CBF), which can construct functions in set $\mathbb{C}$; {\bf (b)} the quantum Bernoulli factories with different quantum processors. They first evolve the $p$-quoins ($|\psi_p\rangle$) with supported operations, and transform the result states into classical probabilities through the measurement which respectively constitute constructible sets denoted by $\mathbb{Q}_1^1$, $\mathbb{Q}^1$ or $\mathbb{Q}$, where the superscript and subscript respectively represent that the constructible states and the unitary operations in the quantum processor are restricted in 1 qubit. The whole QBF processes can construct coins from set $\mathbb{Q}_1^1\mathbb{C}$, $\mathbb{Q}^1\mathbb{C}$ and $\mathbb{Q}\mathbb{C}$ respectively. {\bf (c)} The relationship between the constructible sets. We find the equality among the $\mathbb{QC}$ sets. In summary, we have $\mathbb{Q}_1^1\subseteq\mathbb{Q}^1\subseteq\mathbb{Q}\subsetneq\mathbb{Q}_1^1\mathbb{C}=\mathbb{Q}^1\mathbb{C}=\mathbb{QC}.$}}
  \label{Fig:framework}
\end{figure*}

Basically, the process in QBF can be divided into two phases, the quantum state evolution and the classical coin tossing, and the quantum measurements act as the bridge linking the two phases. The capability of the quantum processor is a key factor to the performance of the QBF. 
For the CBF, there exist no quantum operations, as shown in Fig.~\ref{Fig:framework}{\color{red}{(a)}}. The gradually release of the capabilities of the quantum processors raises three types of QBF, as shown in Fig.~\ref{Fig:framework}{\color{red}{(b)}}.

The first type is the QBF that supports only single qubit unitary operations for quantum state evolution. As has been proved that to cover the whole range of constructible coins of the QBF, the only single-qubit unitary operation used is in the following form~\cite{Dale2015}.
\begin{equation}\label{EQ:UA}
U_a=\left[\begin{matrix}
\sqrt{a} & \sqrt{1-a}\\
\sqrt{1-a} & -\sqrt{a}
\end{matrix}
\right],
\end{equation}
where $a$ is a real number. The functions constructed from the quantum evolution are in a fixed format of $f_a(p)=\left|\sqrt{a(1-p)}-\sqrt{p(1-a)}\right|^2$, which form a set denoted by $\mathbb{Q}_1^1$. In this notation, the `1' in the superscript represents that the output states is limited in single-qubit, and the `1' in the subscript indicates that the unitary operations are restricted in single-qubit. For $a\in(0,1)$, $f_a(p)$ is not classically constructible because $f_a(p)$ reaches 0 when $p=a$. Then, if associated with the further classical processing, it can reach a result set $\mathbb{Q}_1^1\mathbb{C}$ that is strictly larger than that in the CBF.

For the second type, the QBF is enhanced by allowing arbitrary quantum unitary operations for state evolution, but only constructing single-qubit states $|\psi_o\rangle=k_0(p)|0\rangle+k_1(p)|1\rangle$ to generate $\left|k_0(p)\right|^2$-coins for further classical processing. During the process, the quantumly constructed coins form a set denoted by $\mathbb{Q}^1$, and finally the classical processing can reach a set labelled as $\mathbb{Q}^1\mathbb{C}$. Since it has been proved that $\mathbb{Q}^1\subsetneq\mathbb{Q}_1^1\mathbb{C}$~\cite{Jiang2018}, the quantum processor in the type-2 QBF can be completely replaced by a type-1 QBF, while the following duplicated classical processing can not offer new results. This means that the type-2 QBF is structurally equivalent to a type-1 QBF followed by a classical coin processor (see supplementary for details). This obviously indicates that they are equivalent in terms of the range of constructible coins, i.e. $\mathbb{Q}_1^1\mathbb{C}=\mathbb{Q}_1\mathbb{C}$. The enhancement on the quantum processor mainly accelerate the construction for some specific functions~\cite{Dale2015,Patel2019}.


The type-3 QBF uses a quantum processor which supports arbitrary quantum unitary operations and measurements, and generates an $n$-qubit Bernoulli state (see equation ~(\ref{EQ:REWRITTEN})). Then we can apply joint measurement on this state, and  construct a classical coin with success probability of
\begin{equation}\label{eq:qp}
q(p)=\frac{\sum_{j\in\mathbb{H}}\left|h_j(p)\right|^2}{\sum_{i\in\mathbb{B}}\left|h_i(p)\right|^2},
\end{equation}
where $\mathbb{B}$ is the set of chosen bases of measurement corresponding to full probability, and $\mathbb{H}\subseteq \mathbb{B}$ is the set of bases chosen for a head output. The results obtained beyond set $\mathbb{B}$ are discarded. Here we only consider the measurement in the computational basis, since other bases can be converted into computational basis through proper unitary operations. The joint measurement on multi-qubit states allows a wider range of quantumly constructible coins, which forms the set denoted as $\mathbb{Q}$, and the complete set after classical processing is denoted as $\mathbb{Q}\mathbb{C}$. The proposed experiment in ref.~\cite{Patel2019} for $4p(1-p)$-coin is an instance of this strategy, which optimized the experimental realization by directly applying Bell measurement on a 2-qubit state $|\psi_p\rangle|\psi_p\rangle$ where $|\psi_p\rangle = \sqrt{p}|0\rangle+\sqrt{1-p}|1\rangle$. Similarly, we can prove that $\mathbb{Q}\subsetneq \mathbb{Q}_1^1\mathbb{C}$, which further indicates the equality between $\mathbb{Q}_1^1\mathbb{C}$ and $\mathbb{Q}\mathbb{C}$. We place the complete proof in the supplementary methods. In summary, the relationship of the constructible sets can be described by
\begin{equation}
  \mathbb{Q}_1^1\subseteq\mathbb{Q}^1\subseteq\mathbb{Q}\subsetneq\mathbb{Q}_1^1\mathbb{C}=\mathbb{Q}^1\mathbb{C}=\mathbb{QC},
\end{equation}
as shown in Fig.~\ref{Fig:framework}{\color{red}{(c)}}.

\section{Experimental demonstration}

\begin{figure*}[t]
  \centering
  \includegraphics[width=\textwidth]{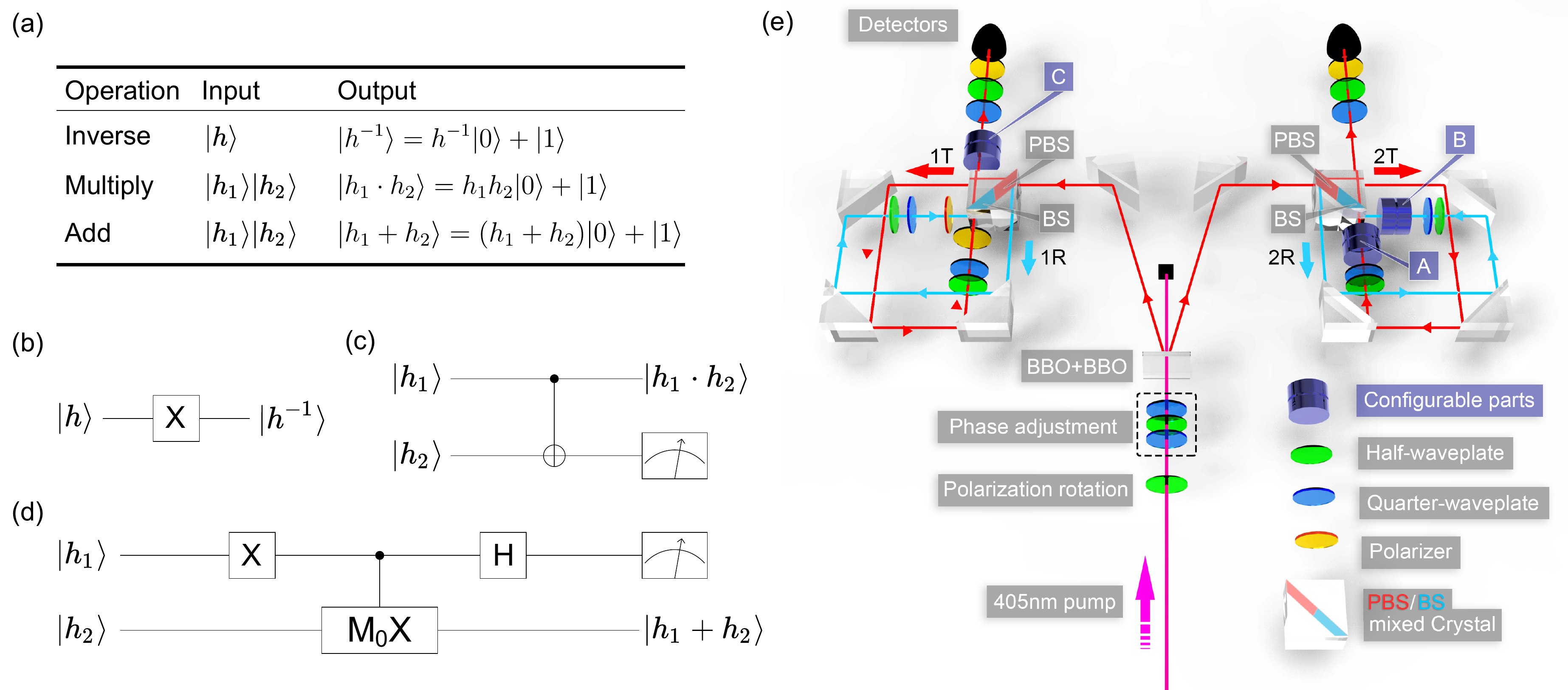}\\
  \caption{\footnotesize{{\bf The experimental proposal.} {\bf (a)} The set of basic operations corresponding to field $\mathbb{M}$. The $h_1$ and $h_2$ are parameters of the input Bernoulli states, which can be constant numbers or functions of $p$, and have been assigned values in the realization. Note that the states are not normalized for simplicity. {\bf (b)} Circuits for the inverse operation; {\bf (c)} Circuits for the multiply operation. It requires two qubits and a C-NOT gate on them; {\bf (d)} Circuits for the add operation. The control-M$_0$X gate is obtained through equation~(\ref{Eq:CM0X}). {\bf (e)} The experimental principle. The main part of the experimental setup is a configurable two-qubit gate implemented with two displaced Sagnac interferometers. The whole process is activated by a pair of entanglement photons produced through type-I SPDC process. $A$, $B$ and $C$ are configurable parts, representing different combinations of optical elements for different operations.}}\label{Fig:ExperimentalProposal}
\end{figure*}

In the experimental demonstration, the hardness concentrates on the implementation of the quantum processor. Without losing generality, we implement the quantum processor of the type-2 QBF, which is the special case of the type-3 QBF and is also an important part of the quantum state generation in the type-3 QBF. The goal of this experiment is to demonstrate the basic operations for manipulating the relative amplitude of an arbitrary given state, 
as shown in Fig.~\ref{Fig:ExperimentalProposal}{\color{red}{(a)}}.

We design the quantum circuits for these operations, as shown in Fig.~\ref{Fig:ExperimentalProposal}{\color{red}{(b-d)}}. Since the inversion requires only a Pauli X gate, we focus on the realization of the multiply and add operations. The multiply operation can be implemented with two qubits and a C-NOT gate. The simplified circuit for add operation uses a C-M$_0$X gate, which is implemented by adding control to a group of gate operations~\cite{Zhou2011, Qiang2016}
\begin{equation}
\label{Eq:CM0X}
{\rm C\text{-}M}_0{\rm X} = {\rm M}_0\otimes {\rm I} + {\rm M}_1\otimes ({\rm M}_0\cdot {\rm X}),
\end{equation}
where ${\rm M}_0$(${\rm M}_1$) is the projection operator corresponding to $|0\rangle$($|1\rangle$). Note that these circuits apply for arbitrary given states.

We built a configurable two-qubit photonic processor, as shown in Fig.~\ref{Fig:ExperimentalProposal}{\color{red}{(e)}}. In the setup, entangled photons are generated through the type-I spontaneous parametric down-conversion (SPDC) process by focusing a diagonally polarized continuous-wave laser beam with central wavelength of $405nm$ onto two orthogonal BBO crystals, generating state $|\psi_0\rangle=(|HH\rangle+|VV\rangle)/\sqrt{2}$, where $|H\rangle$ and $|V\rangle$ represent the horizontal polarization state and vertical polarization state respectively. Then the entangled photons are injected into two {\em Sagnac} structures. Within the Sagnac loops, the entangled photons are converted to be spatially entangled through the PBS part (the red reflecting surface) of the PBS/BS mixed crystal. In each spatial path, a half-waveplate and a quarter-waveplate are used for encoding the input state
\begin{equation}
|\psi_{in}\rangle=\frac{1}{\sqrt{2}}(|h_1\rangle_{1T}|h_2\rangle_{2T}+|h_1\rangle_{1R}|h_2\rangle_{2R}),
\end{equation}
where the $h_1$ and $h_2$ are parameters of the input Bernoulli states, which have been assigned specific values in the realization. The four spacial modes pass through different optical elements. The polarizers in the ``1T'' and ``1R'' paths are fixed to be horizontal and vertical respectively. By placing different combinations of optical elements in the configurable parts labeled as $A$ and $B$ in the logic, we can implement different two-qubit operations. After being mixed in the BS part of the mixed crystal (the reflecting surface marked blue), and further operated by the configurable parts $C$, the state becomes
\begin{equation}
|\psi_o\rangle=(C \otimes {\rm I})({\rm M}_0\otimes A+{\rm M}_1\otimes B)|h_1\rangle|h_2\rangle.
\end{equation}
The state identification is done through a polarizer associated with a quarter-waveplate and a half-waveplate, and another polarizer is used for post-selection. At last, photons are filtered with two $3nm$ band filters.

For the multiply operation, the photonic logic is configured as a C-NOT gate.
Specifically, the parts $A$ and $C$ are configured as identity gates, part $B$ is configured as a Pauli X gate. We then measure the second qubit. If we get $|H\rangle$, the remaining qubit collapsed to $|\psi_\times\rangle = h_1h_2|H\rangle+|V\rangle$. Note that the result state is not normalized. For the add operation, the first X gate can be merged into the state initialization by preparing the initial state to be $|h_1^{-1}\rangle|h_2\rangle$, and then we configure $A$ as an identity gate, $B$ as the combination of an M$_0$ gate and an X gate, and $C$ as a Hadamard gate. We then measure the first qubit, and if we get $|H\rangle$, the remaining qubit will collapse to $|\psi_+\rangle = (h_1+h_2)|H\rangle+|V\rangle$. Our setup employs entangled photons for quantum operations. Recently, a similar experiment was proposed which applied single photons for the realizations of the operations~\cite{Zhan2020}. Compared with our non-unitary realization which directly completes the add operation, they implemented a unitary operation slightly different from the original one from Ref.~\cite{Jiang2018}, and required one more step to complete the add operation.

The key part of the photonic logic is the CNOT gate. We evaluate its fidelity~\cite{Hofmann2005}, and gives that its process fidelity $F_p$ can be bounded as
  $91.40\%\leq F_p\leq 94.16\%$ (see supplementary information for details).
To assess the realized operations, we then take some typical states or random states as input, then measure the fidelities of the output states. The average fidelity of all the states generated by multiply operations and add operations are $95.58\pm4.29\%$ and $96.52\pm4.41\%$ respectively. The results and the parameters of the input states are shown in Fig.~\ref{Fig:FidelityMA}. More detailed data can be found in the supplementary information.

\begin{figure*}[t]
  \centering
  \includegraphics[width=0.93\textwidth]{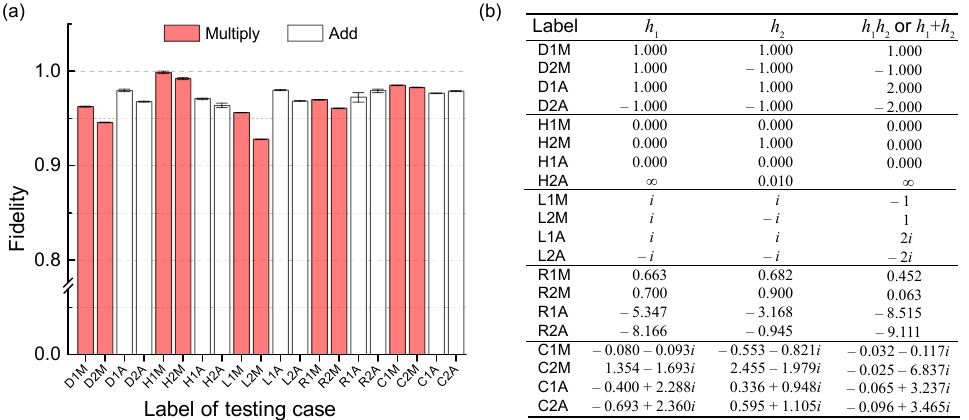}\\
  \caption{\footnotesize{{\bf Results of the add/multiply operations.} (a) The fidelities of the states produced in the multiply and the add operations. The $x$-tick is the label of the test case, which is explained in the table on the right side. (b) The parameters of the input states. The term ``Label'' indicates the type of input states. The initial letter of the label represents the type of the input states, where ``D'' refers to $|D\rangle/|A\rangle$ basis, ``H'' refers to $|H\rangle/|V\rangle$ basis, ``L'' refers to $|L\rangle/|R\rangle$ basis. ``R'' (``C'') refers to states with the parameters chosen to be real (complex) random numbers. $h_1$ and $h_2$ are the parameters of the input states, representing that the input states are $h_1|0\rangle+|1\rangle$ and $h_2|0\rangle+|1\rangle$ respectively (not necessarily normalized). Note that the parameter of $\infty$ indicates the state is exactly $|H\rangle$. The terms ``$h_1h_2$ or $h_1+h_2$'' are the parameters for the produced states.}}\label{Fig:FidelityMA}
\end{figure*}

\section{Experimental Example of the QBF}

Actually, the quantum devices can only approximate the target function, which makes it possible to produce the experimental results with classical protocols. In this case, we show that quantum protocols can also show advantages over the classical ones. We take an example of a $f_c(p)$-coin, where the function $f_c(p)$ is given by
\begin{equation}
  f_c(p)=1-\frac{1}{1+(2p-1)^2}.
\end{equation}
Theoretically, this function is infeasible for classical Bernoulli factory because $f_c(\frac{1}{2}) = 0$, while it can be constructed by directly measuring the Bernoulli state $|f_q(p)\rangle=c_q(p)((2p-1)|0\rangle+|1\rangle)$. We design the circuit for this state by the basic operations, and then simplify it to suit our photonic logic. Specifically, part $A$ and part $B$ are configured so that the sagnac loops act as a C-NOT gate, and part $C$ is consisted of a Hadmard gate and a following Pauli X gate.

\begin{figure}[t]
  \centering
  \includegraphics[width=0.45\textwidth]{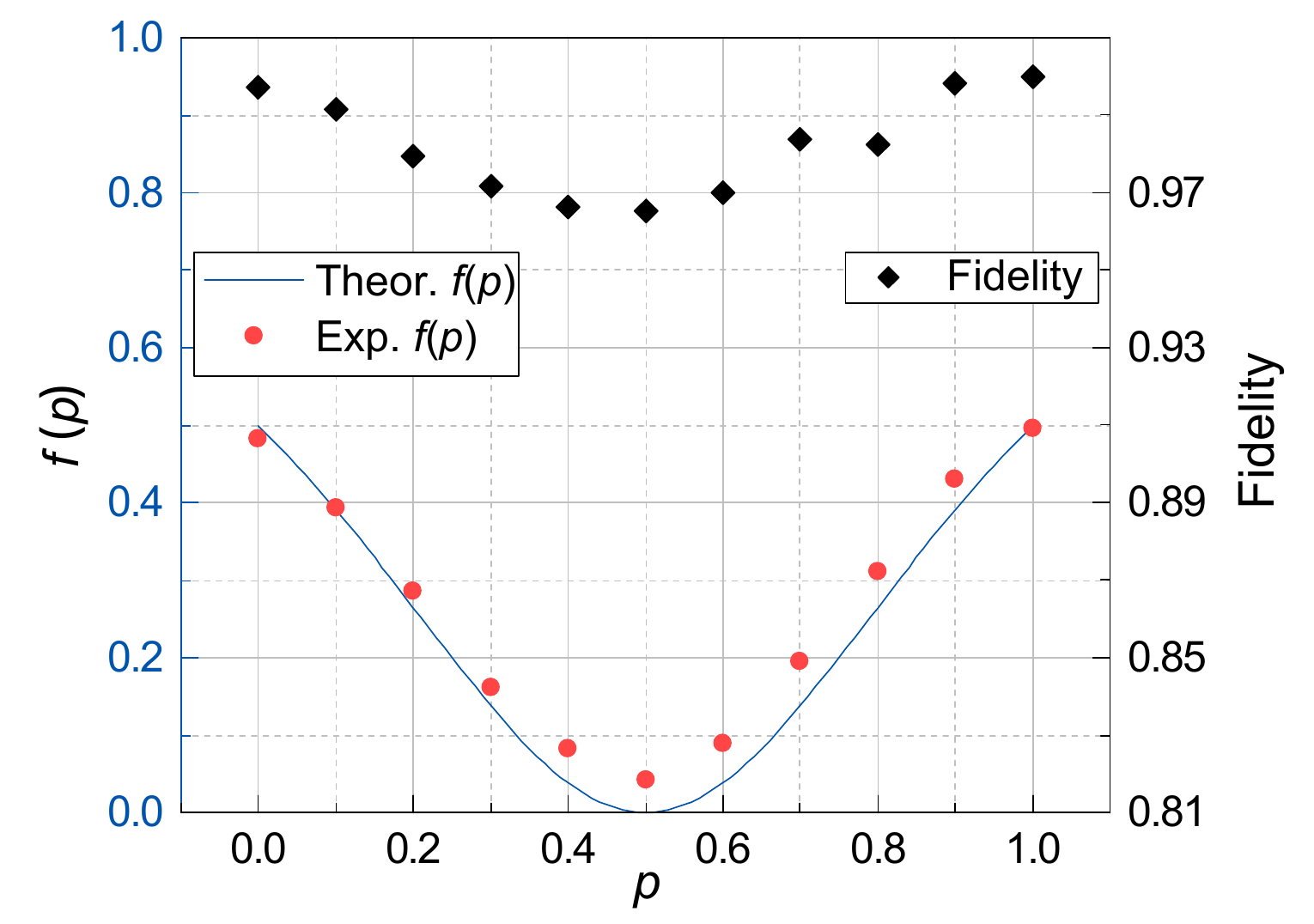}\\
  \caption{\footnotesize{Experimental results of generating $f_c(p)$-coins. The black solid rhombuses are the fidelities of the states corresponding to different value of $p$, where the error bars are too small to be visible (see supplementary information for detailed data). The shaded area is the confidential bound of $f(p)$ estimated according to the fidelities of the produced states. The probabilities (red circles) are obtained through coincidence counts with the total counts accumulated for about 1 minute, and the photon-pair counts recorded ranges from around 400 to about 1,000 according to the different value of $p$.}}\label{Fig:exCoin}
\end{figure}

We measured the fidelities of the outcome states, and then get the success probability of the output coins by measuring the output state in $\sigma_z$ basis. The average fidelity of the states is 98.23\%, and the results agree the theoretical values well, as shown in Fig.~\ref{Fig:exCoin}. The success probability of the construction is $\Pr_c=\left((2p-1)^2+1\right)/16$, which reaches the minimum value of 1/16 when $p=0.5$, this indicates that ideally we need 32 quoins to construct an $f_c$ coin. However owing to the photon loss during the computation, this number would goes up to $\sim54$ (see supplementary information for details).

For comparison, we provide an efficient classical approach to reproduce the experimental results. 
Note that the target function can be rewritten as
\begin{equation}
  f_c(p)=1-\frac{1}{1+(1-g(p))},
\end{equation}
where $g(p)=4p(1-p)$. Obviously, the hardness centers on generating $g(p)$-coins~\cite{Yuan2016,Dale2015}. The generation of $g(p)$-coin can be realized through constructing function $f_1(p)=2p(1-p)$ and $f_2(p)=2p$. For $2p(1-p)$-coins, they can be easily constructed by tossing a $p$-coin twice. However, ideally the $2p$-coin is infeasible for the classical process, but the experimental error results in the subtle rotation of the line. By fitting the experimental results, we found that
it only requires to construct a $1.868p$-coin for $p\in [0,0.5]$ to fit the experimental results. Following ref.~\cite{Huber2016}, the coin consumption is bounded by $9.5C/\epsilon$ where $C = 1.868$ and $\epsilon = 1-C/2$. The overall coin consumption is $N_c\sim1.080\times 10^3$. This protocol consumes 2 orders of magnitude less resources than the related results presented in ref.~\cite{Patel2019} for the linear function, but still consumes about 20-fold more resources than that of the quantum protocol. To the best of our knowledge, this protocol is nearly optimum.

Note that in this case the coin-consumption in the classical protocol scales with the experimental error as $N_c\sim O(\epsilon^{-1})$~\cite{Huber2016}, while in the quantum protocol, the average quoin-consumption is constant after the experiment has been set up. Besides, with the improvement of the experiment accuracy, the corresponding resource consumption of the classical protocol would increase dramatically. In our realization, we presented a more complicated physical realization of the QBF, which corresponds to the basic operations defined in the type-2 QBF, and still demonstrates the advantages over the classical protocol.

\section{Discussion}

In this work, we thoroughly answer whether a $p$-involved state can be implemented from $|\psi_p\rangle$, regardless of the limit on the quantity of qubits. We find the general field structure of the arbitrary Bernoulli states, and show how the multi-qubit states can be used for generating classical probabilities. We further compare three types of quantum Bernoulli factories, and show that the enhancement of the quantum processor can enhance the construction efficiency and reduce the resource consumption.

We experimentally demonstrate the framework of the quantum Bernoulli factory and discuss its advantages in the efficiency and the resource consumption compared with the classical model. Although our quantum realization is not optimal to construct the target function $f_c(p)$, it offers universality for producing a wider range of possible outcomes. The experimental complexity of multi-qubit system introduces more experimental loss, and makes it more sensitive to noise or subtle imperfect settings of the optical elements. But it still shows superiority over the classical protocols. These results may stimulate the potential of the QBF in accelerating the applications where the CBF has already been applied.

\section{acknowledgments}

We appreciate the helpful discussion with other members of QUANTA team. J. W. acknowledges the support from National Natural Science Foundation of China under Grants No. 62061136011 and 61632021. X. S. acknowledges the support from National Natural Science Foundation of China under Grant No. 61832003 and Strategic Priority Research Program of Chinese Academy of Sciences Grant No. XDB28000000. G. T. acknowledges the support from National Natural Science Foundation of China under Grant No. 61801459. J. Z. acknowledges the support from National Natural Science Foundation of China under Grant No. 61872334.




\clearpage
\onecolumngrid

\setcounter{section}{0}
\setcounter{equation}{0}
\setcounter{figure}{0}
\setcounter{table}{0}
\renewcommand{\theequation}{S\arabic{equation}}
\renewcommand{\thefigure}{S\arabic{figure}}
\renewcommand{\thetable}{S\Roman{table}}
\renewcommand{\bibnumfmt}[1]{[S#1]}
\renewcommand{\citenumfont}[1]{S#1}

\begin{center}\bf\large
    Supplementary Materials\\ General Quantum Bernoulli Factory: Framework Analysis and Experiments
\end{center}

\maketitle

\section{The constructibility of the Bernoulli states}

\subsection{Proof of the main theorem}

Our analysis focuses on what quantum states can be constructed from a set of $|\psi_p\rangle=\sqrt{p}|0\rangle+\sqrt{1-p}|1\rangle$. Firstly, we consider the constructibility of the single-qubit Bernoulli states, which is the main theorem of ref.~\cite{Jiang2018S}.

{\bf {\em Theorem}}~\cite{Jiang2018S}. An single-qubit state $$|h(p)\rangle=h(p)|0\rangle+|1\rangle$$ is constructible if and only if every $h(p)$ belongs to $\mathbb{M}$. Note that in this presentation of $|h(p)\rangle$, the coefficient $c(p)$ appeared in eq.~\ref{eq:singlequbit} is ignored. Similarly, we will omit the normalizing coefficient in the following.

{\bf {\em Proof.}} We briefly review its proof. Let $\mathbb{S}$ be the set of $h(p)$ of the constructible single-qubit states, and we need to prove $\mathbb{S}=\mathbb{M}$.

\begin{itemize}

\item {\em Necessity} ($\mathbb{S}\subseteq \mathbb{M}$): The necessity can be easily obtained from the following statement:

  Statement~\cite{Jiang2018S}. For any constructible $|\phi\rangle=\sum_js_j(p)|j\rangle$, the ratio of arbitrary two amplitudes of $|\phi\rangle$ belongs to $\mathbb{M}$.

  It is easy to find that this statement keeps true under any unitary operations and measurement operations. Meanwhile, if this statement is true for $n$ qubit cases, then it is easy to find that it is also true for $n+1$ qubit cases. Based on these two features, this statement can be proved.

\item {\em Sufficiency} ($\mathbb{M}\subseteq \mathbb{S}$): Because set $\mathbb{M}$ is the field generated by the complex field and $\sqrt{\frac{p}{1-p}}$, we need to confirm that $\sqrt{\frac{p}{1-p}}\in\mathbb{S}$, and $\mathbb{S}$ is a field.

    Since we have the initial state as $|\psi_p\rangle=\sqrt{p}|0\rangle+\sqrt{1-p}|1\rangle$, it is obvious that $\sqrt{\frac{p}{1-p}}\in\mathbb{S}$. Also, because all the constant states are constructible, we can know that the complex field is contained in $\mathbb{S}$. The key of the proof resides on that the multiplication, addition and the multiplicative inversion on $p$-involved states are closed in $\mathbb{S}$.

    Suppose that we have $h_1(p),h_2(p)\in\mathbb{S}$, all we have to do is to find unitary operations which realize the three operations.

    \begin{itemize}
      \item {\rm Inverse}: Apply Pauli-$X$ on $|h_1(p)\rangle$, and we get $\frac{1}{h_1(p)}|0\rangle+|1\rangle$, so its multiplicative inverse is in $\mathbb{S}$;
      \item {\rm Multiply}: Apply CNOT on $|h_1(p)h_2(p)\rangle$, and measure the second qubit. If we get $|0\rangle$, the first qubit will be $h_1(p)h_2(p)|0\rangle+|1\rangle$. So $h_1(p)h_2(p)\in\mathbb{S}$, and the set $\mathbb{S}$ is closed under multiplication;
      \item {\rm Add}: Apply $B$ on $|h_1(p)h_2(p)\rangle$, where
      \begin{equation}
        B=\left[\begin{matrix}
          0 &   0   &   0   &   1   \\
          0 &   \frac{1}{\sqrt{2}}   &    \frac{1}{\sqrt{2}}   &   0   \\
          0 &   \frac{1}{\sqrt{2}}   &   -\frac{1}{\sqrt{2}}   &   0   \\
          1 &   0   &   0   &   0   \\
        \end{matrix}\right]
      \end{equation}
      and then measure the first qubit. If we get $|0\rangle$, the rest qubit will be in state $|0\rangle+\frac{h_1(p)+h_2(p)}{\sqrt{2}}|1\rangle$. Then applying the inversion and multiplication, we can know that $h_1(p)+h_2(p)\in\mathbb{S}$, and the set $\mathbb{S}$ is closed under addition.
    \end{itemize}

    So $\mathbb{S}$ is a field, and we complete the proof. $\blacksquare$
\end{itemize}

Now we move to our main theorem, which is a generalization of the theorem in ref.~\cite{Jiang2018S}. Here we recall our main theorem:

{\bf {\em Theorem.}} An $n$-qubit state $$|K(p)\rangle=\sum_{i=0}^{2^n-2}h_i(p)|i\rangle+|2^n-1\rangle$$ is constructible if and only if every $h_i(p), (i=0,1,\cdots,2^n-2)$ belongs to $\mathbb{M}$.

{\bf {\em Proof.}} Let $\mathbb{S}$ be the set of relative amplitudes of any constructible states. Here we prove $\mathbb{S}=\mathbb{M}$. For simplicity, we omit the normalization coefficient $c(p)$ appeared in equation~(\ref{EQ:REWRITTEN}) in the main text.

\begin{itemize}
  \item {\em Necessity} ($\mathbb{S}\subseteq \mathbb{M}$): The necessity can be proved according to the same statement.

  \item {\em Sufficiency} ($\mathbb{M}\subseteq \mathbb{S}$): Suppose we have implemented a state $|K(p)\rangle=\sum_{i=0}^{2^n-2}h_i(p)|i\rangle+|2^n-1\rangle$, and another single-qubit state $|L(p)\rangle=l(p)|0\rangle+|1\rangle$ where $h_i(p)\in \mathbb{S}, (i=0,1,2,...,2^n-2)$ and $l(p)\in \mathbb{S}$. Without loss of generality, we show that the relative amplitude of $|0\rangle$ in $|K(p)\rangle$ can be manipulated without changing other relative amplitudes. The other relative amplitudes can be manipulated in the similar way. The set that $h_0(p)$ belongs to is closed under addition and multiply, and containing multiplicative inverse for each element. Thus the set $\mathbb{S}$ is a field which contains element $\sqrt{\frac{p}{1-p}}$ and complex field. In other words, $\mathbb{S}=\mathbb{M}$. The details are shown as below.
  \begin{itemize}
    \item {\em Inverse}: because $h_0(p)\in \mathbb{S}$, we can implement a single-qubit state $|h_0(p)\rangle=c_0(p)(h_0(p)|0\rangle+|1\rangle)$ according to the theorem of ref.~\cite{Jiang2018S}. We then switch the amplitudes of $|0\rangle|0\rangle^{\otimes n}$ and $|1\rangle|2^n-1\rangle$ of $|h_0(p)\rangle|K(p)\rangle$, and then measure the first qubit. If we get $|0\rangle$, the rest qubits will collapse to
        \begin{equation}
        |K_{h_0^{-1}}(p)\rangle=\frac{1}{h_0(p)}|0\rangle+\sum_{i=1}^{2^n-2}h_i(p)|i\rangle+|2^n-1\rangle.
        \end{equation}
        So $\frac{1}{h_0(p)}\in \mathbb{S}$.
    \item {\em Multiply}: apply an $(n+1)$-qubit unitary operation to switch the amplitudes of $|0\rangle|0\rangle^{\otimes n}$ and $|1\rangle|0\rangle^{\otimes n}$ of $|L(p)\rangle|K(p)\rangle$, and measure the first qubit. If we get $|1\rangle$, the remaining qubits will collapse to
        \begin{equation}
        |K_{M_0}(p)\rangle=h_0(p)l(p)|0\rangle+\sum_{i=1}^{2^n-2}h_i(p)|i\rangle+|2^n-1\rangle.
        \end{equation}
        Thus $h_0(p)l(p)\in \mathbb{S}$.
    \item {\em Add}: it requires an specific $2^{n+1}\times 2^{n+1}$ unitary $B_n$. This unitary looks like an eye matrix, with part of its diagonal similar to a Hadmard matrix
        \begin{equation}\label{EQ:UB}
          B_n=\left[
          \begin{matrix}
          1 &   &   &   &   &   &     \\
            & \ddots &   &   &   &   &     \\
            &   & \ddots &   &   &   &   &     \\
            &   &   & \frac{1}{\sqrt{2}} & \dots & -\frac{1}{\sqrt{2}} &   &      \\
            &   &   & \vdots &   & \vdots &   &      \\
            &   &   & \frac{1}{\sqrt{2}} & \dots & \frac{1}{\sqrt{2}} &   &      \\
            &   &   &   &   &   & \ddots &     \\
            &   &   &   &   &   &   &  1   \\
          \end{matrix}
          \right].
          \end{equation}
          The up left $\frac{1}{\sqrt{2}}$ appears at the $2^n$-th diagonal position, and the bottom right $\frac{1}{\sqrt{2}}$ is in the $(2^n+(k+1))$-th position, where $|k\rangle$ is the basis to conduct the add operation. Without loss of generality, we are going to manipulate the relative amplitude $h_0(p)$ with $k=0$.

          Then we apply $B_n$ on $|L(p)\rangle|K(p)\rangle$, and measure the first qubit. If we get $|1\rangle$, the state will collapse to
          \begin{equation}
          |K_{A_0}(p)\rangle=\frac{h_0(p)+l(p)}{\sqrt{2}}|0\rangle+\sum_{i=1}^{2^n-2}h_i(p)|i\rangle+|2^n-1\rangle.
          \end{equation}
          By multiplying $|K_{A_0}\rangle$ with a single-qubit constant state $|\psi_a\rangle=\sqrt{2}|0\rangle+|1\rangle$ on basis $|0\rangle$ of $|K_{A_0}(p)\rangle$, we can obtain $h_0(p)+l(p)\in \mathbb{S}$.

          \end{itemize}

  Therefore, $\mathbb{S}$ is a field. We initially have access to $|p\rangle$ and arbitrary constant qubits, so the generator of $\mathbb{S}$ contains $\sqrt{\frac{p}{1-p}}$ and the complex field, and we conclude $\mathbb{M}\subseteq \mathbb{S}$. Combining the necessity and sufficiency, we complete our proof.$\blacksquare$
\end{itemize}

By using the operations, we know that each amplitude can be manipulated without changing other amplitudes, i.e. each amplitude can be manipulated independently. To apply the basic operations on different amplitudes, one can just slightly modify the auxiliary states and unitary matrix used, and follow the same procedure.

For example, if we have a 2-qubit state $|K_2\rangle=h_0(p)|00\rangle+h_1(p)|01\rangle+h_2(p)|10\rangle+|11\rangle$, and a single-qubit state $|l(p)\rangle=l(p)|0\rangle+|1\rangle$. We now want to add $l(p)$ onto the relative amplitude of $|1\rangle$ in state $|K_2\rangle$, we first apply unitary $B_2$ on $|l(p)\rangle|K_2\rangle$, where

\begin{equation}
  B_2=\left[\begin{matrix}
            1 & 0 & 0 & 0                   & 0 & 0                     & 0 & 0 \\
            0 & 1 & 0 & 0                   & 0 & 0                     & 0 & 0 \\
            0 & 0 & 1 & 0                   & 0 & 0                     & 0 & 0 \\
            0 & 0 & 0 & \frac{1}{\sqrt{2}}  & 0 & -\frac{1}{\sqrt{2}}   & 0 & 0 \\
            0 & 0 & 0 & 0                   & 1 & 0                     & 0 & 0 \\
            0 & 0 & 0 & \frac{1}{\sqrt{2}}  & 0 & \frac{1}{\sqrt{2}}    & 0 & 0 \\
            0 & 0 & 0 & 0                   & 0 & 0                     & 1 & 0 \\
            0 & 0 & 0 & 0                   & 0 & 0                     & 0 & 1 \\
       \end{matrix}\right],
\end{equation}

and we have

\begin{equation}
\begin{aligned}
    &B_2|l(p)\rangle|K_2\rangle\\
   =&|0\rangle\otimes\left(l(p)h_0(p)|00\rangle+l(p)h_1(p)|01\rangle+l(p)h_2(p)|10\rangle+\frac{l(p)-h_1(p)}{\sqrt{2}}|11\rangle\right)\\
    &+|1\rangle\otimes\left(h_0(p)|00\rangle+\frac{h_1(p)+l(p)}{\sqrt{2}}|01\rangle+h_2(p)|10\rangle+|11\rangle\right).\\
\end{aligned}
\end{equation}

Then, we measure the first qubit. If we get $|1\rangle$, we then obtain
\begin{equation}
  |K_{2A}\rangle=h_0(p)|00\rangle+\frac{h_1(p)+l(p)}{\sqrt{2}}|01\rangle+h_2(p)|10\rangle+|11\rangle.
\end{equation}

Finally, we multiply $\sqrt{2}|0\rangle+|1\rangle$ on the relative amplitude of $|01\rangle$ in state $|K_{2A}\rangle$, we obtain
\begin{equation}
  h_0(p)|000\rangle+(h_1(p)+l(p))|01\rangle+h_2(p)|10\rangle+|11\rangle,
\end{equation}
and the other relative amplitudes are not changed.

\subsection{The algorithm for generating arbitrary constructible $n$-qubit states}

If $n=1$, i.e. we are going to construct a single-qubit Bernoulli factory, we can construct it following the procedure in ref.~\cite{Jiang2018S}.

If $n>1$, our method starts from a constant balanced $n$-qubit state (not necessarily normalized)
\begin{equation}
|\Psi\rangle=\sum_{i=0}^{2^n-1}|i\rangle.
\end{equation}
For $|F_n(p)\rangle=\sum_{i=0}^{2^n-2}f_i(p)|i\rangle+|2^n-1\rangle$ where $f_i(p)\in\mathbb{M}$, we can firstly generate a series of single-qubit states
\begin{equation}
|f_i(p)\rangle=f_i(p)|0\rangle+|1\rangle,~(i=0,1,...,2^n-2),
\end{equation}
and then multiply $|f_i(p)\rangle$ with $|\Psi\rangle$ on the corresponding basis one by one.

\section{The framework analysis of quantum Bernoulli factory}

We divided the QBF into 3 types according to the capabilities of their quantum processors. The type-1 QBF supports only single qubit operations, the type-2 QBF supports arbitrary unitary operations but only produces single-qubit states, and the type-3 QBF provides a universal condition for states construction. Now we provide more details about the analysis of the bound of the constructible sets of these three types of QBFs.

\subsection{Proof of $\mathbb{Q}\subseteq\mathbb{Q}_1^1\mathbb{C}$}

Recall that $\mathbb{Q}_1^1\mathbb{C}$ is the final constructible set of type-1 QBF, and $\mathbb{Q}$ is the set of classical coins that can be generated by measuring a Bernoulli state that contains no less than one qubit.

{\bf \em Definition}~\cite{Dale2015S}. A function $f(p):[0,1]\rightarrow [0,1]$ is simple and poly-bounded (SPB) if and only if it satisfies

(1) $f$ is continuous.

(2) Both $Z={z_i:f(z_i)}=0$ and $W={w_i:f(w_i)}=1$ are finite sets.

(3) $\forall z\in Z$, there exist constants $c,\delta>0$ and integer $k<\infty$ such that
\begin{equation}
c(p-z)^{2k}\leq f(p),\forall p\in [z-\delta, z+\delta].
\end{equation}

(4) $\forall w\in W$, there exist constants $c,\delta>0$ and integer $k<\infty$ such that
\begin{equation}
1-c(p-w)^{2k}\geq f(p),\forall p\in [w-\delta, w+\delta].
\end{equation}

{\bf {\em Lemma} 1}~\cite{Dale2015S}. A function is constructible in quantum Bernoulli factory with $|\psi_p\rangle=\sqrt{p}|0\rangle+\sqrt{1-p}|1\rangle$ and a set of single-qubit unitary operations if and only if $f$ satisfies SPB conditions.

{\bf {\em Lemma} 2}~\cite{Jiang2018S}. Let $T(x_1, x_2, x_3):\mathbb{R}^3\rightarrow \mathbb{R}$ be a multivariate polynomial of $x_1$, $x_2$ and $x_3$. Suppose $T(p,\sqrt{p},\sqrt{1-p})$ is not a zero function. If $T(z,\sqrt{z},\sqrt{1-z})=0$ for some $z\in[0,1]$. Then there exist a real number $\delta$, an integer $k$ and a function $m(p)$ which is continuous in $[z-\delta, z+\delta]$, such that $T(p,\sqrt{p},\sqrt{1-p})=(p-z)^{\frac{1}{2}k}m(p)$ and $m(z)\neq 0$.

{\bf {\em Proof}}. The proof of $\mathbb{Q}\subseteq\mathbb{Q}_1^1\mathbb{C}$ is similar to the proof of $\mathbb{Q}^1\subseteq\mathbb{Q}_1^1\mathbb{C}$. The difference is concentrated on dealing with the continuity of $f$.

For the classical coins from $\mathbb{Q}$
\begin{equation}
q(p)=\frac{\sum_{j\in\mathbb{H}}\left|h_j(p)\right|^2}{\sum_{i\in\mathbb{B}}\left|h_i(p)\right|^2},
\end{equation}
where $\mathbb{B}$ is the set of bases remained within the post selection of the measurement, and $\mathbb{H}\subseteq \mathbb{B}$ is the set of bases chosen for a head output. According to the main theorem, there exist a series of complex multivariate polynomials $R_i(x_1,x_2,x_3)$, such that
\begin{equation}
f(p)=\frac{\sum_{j\in\mathbb{H}}\left|R_j(p, \sqrt{p}, \sqrt{1-p})\right|^2}{\sum_{i\in\mathbb{B}}\left|R_i(p, \sqrt{p}, \sqrt{1-p})\right|^2}.
\end{equation}

For arbitrary $f(p)\in \mathbb{Q}$, we prove that $f(p)$ satisfies the SPB conditions.

\begin{itemize}
  \item (1) $f(p)$ is continuous.
    The only issue to consider about is that there might be some strange points that belong to both the zeros of the dominator and numerator. For example, we can construct a state
    \begin{equation}
    |\phi\rangle=(1-2p)|0\rangle+(1-2p)^2(1-3p)|1\rangle+(1-2p)(1-4p)^2|2\rangle+|3\rangle.
    \end{equation}
    Note that this state is not normalized. Then we can obtain the classical function
    \begin{equation}
    s(p)=\frac{(1-2p)^2+(1-2p)^4(1-3p)^2}{(1-2p)^2+(1-2p)^4(1-3p)^2+(1-2p)^2(1-4p)^4}.
    \end{equation}
    via measurement if the outcome is $|0\rangle$ or $|1\rangle$ on condition of obtaining $|0\rangle$, $|1\rangle$ or $|2\rangle$. This function is continuous on $[0,0.5)\bigcup(0.5,1]$, and for other $p\in[0,1]$, it behaves exactly the same with
    \begin{equation}\label{eq:NoCommonZeros}
    s_e(p)=\frac{1+(1-2p)^2(1-3p)^2}{1+(1-2p)^2(1-3p)^2+(1-4p)^4}.
    \end{equation}
    Fortunately, we can handle this exception by using a small trick. Let $f_e(p)$ be the function after extracting factors involving the common zeros, so that there is no common zero between the numerator and dominator in $f_e(p)$ (such as $s_e(p)$ in equation~(\ref{eq:NoCommonZeros})). We denote the zeros of $R_i$ as $Z_i$. It is easy to show that $f(p)$ is continuous in $[0,1]-\bigcap_{i\in\mathbb{B}}Z_i$, and obviously, $f_e(p)$ is continuous in $[0,1]$, and therefore $f_e(p)$ satisfies the SPB conditions, i.e. $f_e(p)\in\mathbb{Q}_1^1\mathbb{C}$.

    Since $f_e(p):[0,1]\rightarrow [0,1]$ is constructible, it is of course that $f_e(p)$ is constructible when we limit the range of $p$, i.e. $f_e(p)$ is constructible for $p\in [0,1]-\bigcap_{i\in\mathbb{B}}Z_i$. Therefore, we can simulate $f_e(p)$ for $p\in [0,1]-\bigcap_{i\in\mathbb{B}}Z_i$, which is equivalent to simulate $f(p)$. In other words, we can extended $f(p)$ to $[0,1]$, which is exactly $f_e(p)$. In this way, we can handle the continuity of the functions.
  \item (2) Both $Z={z_i:f(z_i)}=0$ and $W={w_i:f(w_i)}=1$ are finite sets.
    Because $R_i(x_1, x_2, x_3)$ are multivariate polynomials, so $|R_i(p, \sqrt{p}, \sqrt{1-p})|$ is bounded when $p\in[0,1]$, and has finite zeros in [0,1]. We can then find out that the set of $Z$ is
    \begin{equation}
    Z=\left(\bigcap_{j\in\mathbb{H}}Z_j\cap [0,1]\right) - \bigcap_{i\in\mathbb{B}}Z_i,
    \end{equation}
    and obviously $Z$ is finite. Similarly, $W=\left(\bigcap_{i\notin \mathbb{H}}Z_i\right)\cap [0,1]$ is finite. In summary, both $Z$ and $W$ are finite. It is worth noting that at the breaking points of $f$ (i.e. the joint set of all $Z_i$ for $i\in\mathbb{B}$), the function $f$ can be extended to be a continuous one with the inserted value given by equation~(\ref{eq:NoCommonZeros}).
  \item (3) $\forall z\in Z$, there exist constants $c,\delta>0$ and integer $k<\infty$ such that
    \begin{equation}
    c(p-z)^{2k}\leq f(p),\forall p\in [z-\delta, z+\delta].
    \end{equation}
    This can be easily checked using Lemma 2.
  \item (4) $\forall w\in W$, there exist constants $c,\delta>0$ and integer $k<\infty$ such that
    \begin{equation}
    1-c(p-w)^{2k}\geq f(p),\forall p\in [w-\delta, w+\delta].
    \end{equation}
    This just requires to have $1-f(p)\geq c(p-w)^{2k}$. Note that $w$ is then one of the zeros of $1-f(p)$. Therefore the satisfiability of this condition can be similarly obtained through Lemma 2.$\blacksquare$
\end{itemize}

\subsection{The equality of the $\mathbb{QC}$ sets}

\begin{figure}[!htb]
  \centering
  \includegraphics[width=0.85\textwidth]{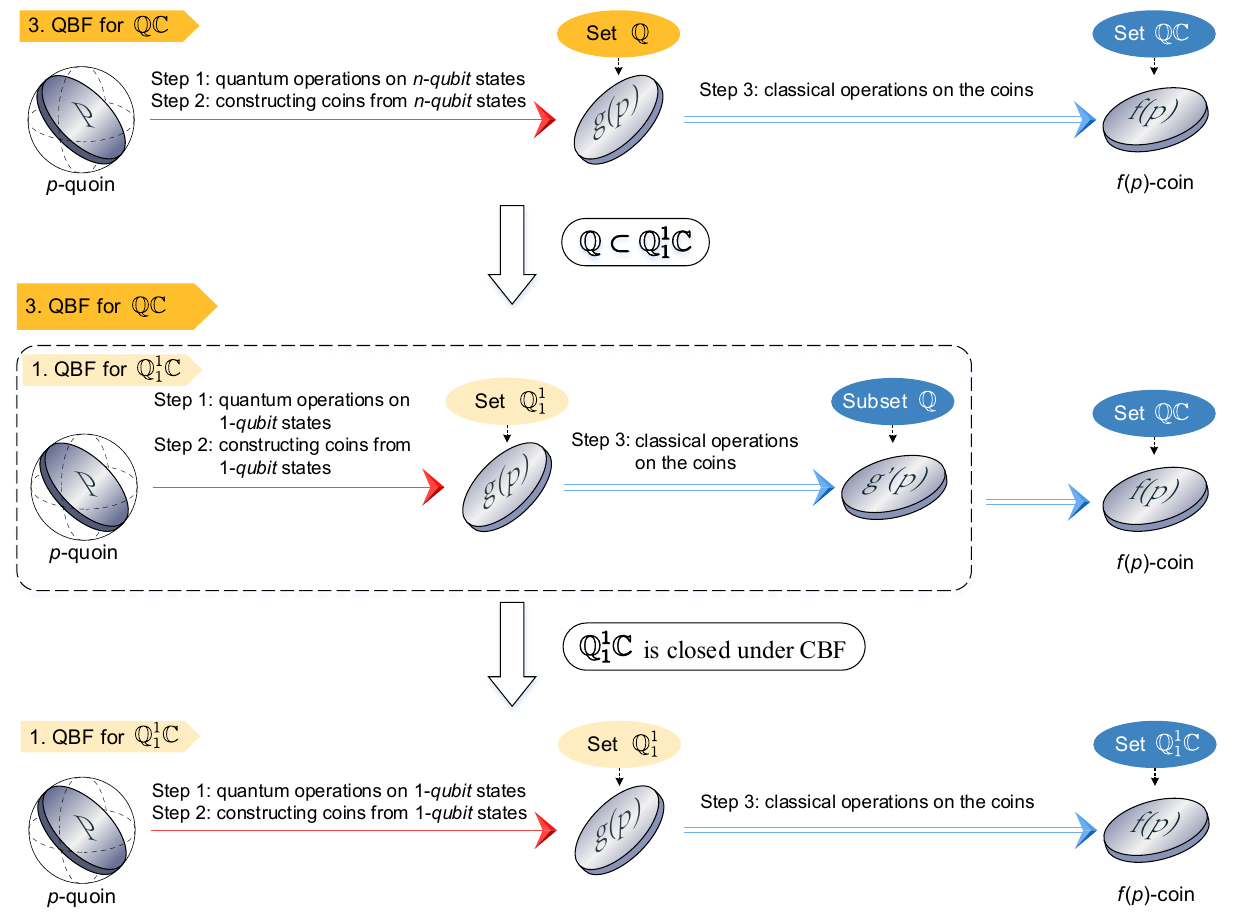}\\
  \caption{\footnotesize{{The capability of QBFs in different types are the same.}}}\label{Fig:EQQC}
\end{figure}

We provide an illustrative proof for this result, as shown in Fig.~\ref{Fig:EQQC}. We firstly show that $\mathbb{QC}=\mathbb{Q}_1^1\mathbb{C}$. The quantum operations for type-3 QBF can reach a set denoted as $\mathbb{Q}$, and with the above results, we know that $\mathbb{Q}\subset\mathbb{Q}_1^1\mathbb{C}$. Therefore, we can just replace the quantum processor of type-3 QBF with a complete type-1 QBF. Then, because of that $\mathbb{Q}_1^1\mathbb{C}$ is closed under the classical processing, the whole process is equivalent with a standard type-1 QBF, and we obtain the result that $\mathbb{QC}=\mathbb{Q}_1^1\mathbb{C}$. Owing to the same reason, we also have that $\mathbb{Q}^1\mathbb{C}=\mathbb{Q}_1^1\mathbb{C}$.

\subsection{The QBF Protocols for $g(p)=4p(1-p)$-coin}
\label{Sec:QBFProtocol}

Here we show protocols in view of the framework of QBF for constructing function $g(p)=4p(1-p)$.
The classical function $g(p)$ is an important function, that servers as the core elements for many other functions. In ref.~\cite{Yuan2016S}, this function is experimentally constructed utilizing quantum coherence, and the results showed that quantum entanglement is not necessary for the construction. However, it has been experimentally shown that when utilizing quantum entanglement, the construction efficiency can be greatly enhanced, along with the reduction of resource consumption by orders of magnitudes compared with the cases where only single-qubit operations are used~\cite{Patel2019S}.

{\bf The type-1 QBF protocol}. The first protocol uses single-qubit operations, which has the following procedures.
\begin{enumerate}
  \item (Quantum processing) generate a $p$-coin, which is done by directly measuring $|\psi_p\rangle$ in $\sigma_z$ basis. Recall that a $f(p)$-coin is a classical coin with probability $f(p)$ to output a head, and $1-f(p)$ to output a tail;
  \item (Quantum processing) generate a $q$-coin, where $q=\frac{1+2\sqrt{p(1-p)}}{2}$. This coin can be constructed by applying Hadmard gate on $|\psi_p\rangle$ and then measure it in $\sigma_z$ basis, or directly measuring $|\psi_p\rangle$ in $D/A$ basis;
  \item (Classical processing) construct an $m$-coin from $p$-coins, where $m=2p(1-p)$. This is a purely classical process by tossing the $p$-coin twice. Similarly, one can construct an $n$-coin where $n=2q(1-q)=1/2-2p(1-p)$;
  \item (Classical processing) construct a $s$-coin from $m$-coins, where $s=m/(m+1)$. Toss the $m$-coin twice, if the first toss is tail, then output tail; otherwise if the second toss is tail, output head; otherwise if both tosses are head, repeat this step.  Similarly, construct a $t$-coin from $n$-coins, where $t=n/(n+1)$;
  \item (Classical processing) construct a $g(p)=4p(1-p)$-coin by toss a $s$-coin and then toss a $t$-coin. If the first toss is head, and the second toss is tail, then output head; if the first toss is tail and the second toss is head, then output tail; otherwise repeat this step.
\end{enumerate}
The quantum state evolution in these procedures involves only single-qubit operation, which can be easily done on our photonic processor.

{\bf The type-2 QBF protocol}. The second protocol for $g(p)$-coin is finished in type-2 QBF, where arbitrary quantum operations can be applied for generating a single-qubit Bernoulli state, which is then measured to produce a classical coin for classical processing. Note that in this case, the joint-measurement is not allowed because this kind of operations are equivalent to multi-particle operations. Therefore the Bell-measurement protocol metioned in ref.~\cite{Dale2015S} is beyond the capability of type-2 QBF. In type-2 QBF protocol, the $g(p)$-coin can be generated by solely quantum operations and no more classical processes are required, because there exists a single-qubit state for the $g(p)$-coin:
\begin{equation}
|\psi_g\rangle=\sqrt{4p(1-p)}|0\rangle+(2p-1)|1\rangle.
\end{equation}
Measuring $|\psi_g\rangle$ in $\sigma_z$ basis can directly obtain the $g(p)$-coin. Assuring the existence of the corresponding Bernoulli state, we can then optimize the circuit that constructs $|\psi_g\rangle$, as shown in the following procedures.
\begin{enumerate}
  \item (Quantum processing) apply CNOT on $|\psi_p\rangle|\psi_p\rangle$, resulting in
  \begin{equation}
    |\psi_1\rangle=p|00\rangle+\sqrt{p(1-p)}|01\rangle+(1-p)|10\rangle+\sqrt{p(1-p)}|11\rangle.
  \end{equation}
  \item (Quantum processing) apply Hadmard operation on the first qubit to obtain
  \begin{equation}\label{eq:psi2}
    |\psi_2\rangle=\frac{1}{\sqrt{2}}|00\rangle+\sqrt{2p(1-p)}|01\rangle+\frac{(2p-1)}{\sqrt{2}}|10\rangle.
  \end{equation}
  \item (Quantum processing) apply CNOT on $|\psi_2\rangle$, and post-selecting the second qubit in $|1\rangle$ basis will give
  \begin{equation}
    |\psi_g\rangle=\sqrt{4p(1-p)}|0\rangle+(2p-1)|1\rangle.
  \end{equation}
  \item (Quantum processing) measure $|\psi_g\rangle$ will result in the target function.
\end{enumerate}

{\bf The type-3 QBF protocol}. The third protocol for $g(p)$-coin is generating a multi-qubit Bernoulli states for probability measurement. In fact, we just need to prepare $|\Psi_{ini}\rangle=|\psi_p\rangle|\psi_p\rangle$, and choose the set of measuring base as $\mathbb{B}=\{|\Psi^{+}\rangle, |\Phi^{-}\rangle$ and $\mathbb{H} = \{|\Psi^+\rangle\}$, where $|\Psi^{\pm}\rangle=(|01\rangle\pm|10\rangle)/\sqrt{2}, |\Phi^{\pm}\rangle=(|00\rangle\pm|11\rangle)/\sqrt{2}$. Note that in this case what $\mathbb{B}$ and $\mathbb{H}$ contain are not the computational bases, and eq.~\ref{eq:qp} obtained can be described by:
\begin{equation}
  q(p)=\frac{\sum_{|h\rangle\in\mathbb{H}} \langle\Psi_{ini}|h\rangle\langle h|\Psi_{ini}\rangle}{\sum_{|b\rangle\in\mathbb{B}} \langle\Psi_{ini}|b\rangle\langle b|\Psi_{ini}\rangle}=\frac{\sum_{|h\rangle\in\mathbb{H}} \left|\langle\Psi_{ini}|h\rangle\right|^2}{\sum_{|b\rangle\in\mathbb{B}} \left|\langle\Psi_{ini}|b\rangle\right|^2}.
\end{equation}

To show this more clearly, we finish this process in computational bases. The first two steps are the same with the second protocol, then we directly measure the probability from this 2-qubit state $|\psi_2\rangle$ (see eq.~\ref{eq:psi2}), and $g(p)$ is the probability of obtaining $|01\rangle$ on condition of obtaining $|01\rangle$ and $|10\rangle$
\begin{equation}
g(p)=\frac{\left|h_{|01\rangle}(p)\right|^2}{\sum_{i\in\{|01\rangle,|10\rangle\}}\left|h_{i}(p)\right|^2}
=\frac{\left|\sqrt{2p(1-p)}\right|^2}{\left|\sqrt{2p(1-p)}\right|^2+\left|(2p-1)/\sqrt{2}\right|^2}=4p(1-p).
\end{equation}

Obviously, the type-3 QBF protocol is the most efficient. This indicates that with the enhancement of the quantum processors, the whole process can be accelerated.

\subsection{The definitions of the sets}

During the analysis, there appears various sets of constructible functions. Here we list the definitions of the sets.

\begin{itemize}
  \item Set $\mathbb{M}$ and set $\mathbb{S}$

  The set $\mathbb{M}$ is the field generated from $\sqrt{\frac{p}{1-p}}$ and the complex field, where $p$ is an unknown parameter. The formula form of set $\mathbb{M}$ is shown in eq.~\ref{eq:setM}. $\mathbb{S}$ is the set of all the possible relative amplitudes of the constructible states, which is equal to set $\mathbb{M}$.

  \item Set $\mathbb{C}$

  This set contains all constructible functions of classical Bernoulli factory. The characterization of this set can be described by the three conditions~\cite{Keane1994S}: (1) The function should be continuous on its domain; (2) The function should not reach 0 or 1 within its domain; (3) The function should not approach 0 or 1 exponentially fast near any edge of its domain.

  \item Set $\mathbb{Q}_1^1$

  It is the set of classical functions constructed by the quantum processor that evolves $|\psi_p\rangle$ using a specific type of unitary (eq.~\ref{EQ:UA} in main text), and therefore this set can be written as
  \begin{equation}
  \mathbb{Q}_1^1=\left\{f_a(p)|a\in(0,1)\right\}.
  \end{equation}
  where $f_a(p)=\left|\sqrt{a(1-p)}-\sqrt{p(1-a)}\right|^2$. These functions are those used in the method of ref.~\cite{Dale2015S}.

  \item Set $\mathbb{Q}^1$

  This set contains the constructible functions by using the quantum processor embedded within the type-2 QBF to construct a single-qubit Bernoulli state, and then measure this state to obtain a classical result. Specifically,

  \begin{equation}
  \mathbb{Q}^1=\left\{\left|k_0(p)\right|^2\Big| \frac{k_0(p)}{k_1(p)}\in\mathbb{M}\right\}.
  \end{equation}
  where $k_0(p)$ and $k_1(p)$ are functions of $p$ and satisfy $|k_0(p)|^2+|k_1(p)|^2=1$.

  \item Set $\mathbb{Q}$

  This set contains the constructible functions by the quantum processor of type-3 QBF. This is the general case for set $\mathbb{Q}^1$ where the states generated are free of restrictions on the number of qubits. Specifically, we generate an $n$-qubit Bernoulli state $|K(p)\rangle=c(p)\left(\sum_{i=0}^{2^n-2}h_i(p)|i\rangle + |2^n-1\rangle\right)$, and obtain a function in set
  \begin{equation}
  \mathbb{Q}=\left\{\frac{\sum_{j\in\mathbb{H}}|h_j(p)|^2}{\sum_{i\in\mathbb{B}}|h_i(p)|^2}\Big|h_i(p)\in\mathbb{M}\right\},
  \end{equation}
  where each $h_i(p)$ is the relative amplitude of $|K(p)\rangle$, $\mathbb{B}$ is the set of bases corresponding to the full probability in the measurement, and $\mathbb{H}\subseteq\mathbb{B}$ is the set of bases chosen for a head output.

  \item Set $\mathbb{Q}_1^1\mathbb{C}$

  This set is the constructible functions of the type-1 QBF obtained by feeding functions in $\mathbb{Q}_1^1$ into the classical processing. This set is bounded by three conditions: (1) the function should be continuous in its domain. (2) this function should reach 0 or 1 in its domain for finite times. (3) The function should not approach 0 or 1 exponentially fast on any edge of its domain. The detailed characterization of this set is analyzed in~\cite{Dale2015S}.

  \item Set $\mathbb{Q}^1\mathbb{C}$

  This set contains the constructible functions of the type-2 QBF, obtained by feeding functions in $\mathbb{Q}^1$ into the classical processing. We find that this set is equal to $\mathbb{Q}_1^1\mathbb{C}$.

  \item Set $\mathbb{Q}\mathbb{C}$

  This set contains the constructible functions of the type-3 QBF, obtained by feeding functions in $\mathbb{Q}$ into the classical processing. We find that this set is equal to $\mathbb{Q}_1^1\mathbb{C}$.

\end{itemize}

\section{Experimental demonstration of the framework of QBF}

FIG.~\ref{Fig:ExperimentalProposal}{\color{red}{(e)}} in the main text shows the experimental proposal. The entangled photons are obtained through the type-1 SPDC process. Before being injected into the two Sagnac interferometers, the photons are in the following state:
\begin{equation}
|\phi\rangle=\frac{1}{\sqrt{2}}\left(|00\rangle+|11\rangle\right),
\end{equation}
where $|0\rangle$ and $|1\rangle$ represent the horizontal and vertical polarization respectively. Then each of the two photons goes into a Sagnac interferometer, which consists of a PBS/BS mixed crystal and three prisms. The PBS part of the cube converts the polarization-entanglement to the spatial entanglement, so that the state becomes:
\begin{equation}
|\phi_{spatial}\rangle=\frac{1}{\sqrt{2}}\left(|0\rangle_{1T}|0\rangle_{2T}+|1\rangle_{1R}|1\rangle_{2T}\right),
\end{equation}
where $1T$, $2T$, $1R$ and $2R$ represent different spatial modes labelled in FIG.~\ref{Fig:ExperimentalProposal}{\color{red}{(e)}} (main text). Four groups of wave-plates (including one HWP and one QWP), denoted by $T_{1H}$, $T_{1V}$, $T_{2H}$ and $T_{2V}$ (the labels are not marked in the figure) are placed in each path. These waveplates work on the polarization of the four spacial modes, and prepare the states into
\begin{equation}\label{Eq:initialState}
\left\{
\begin{aligned}
T_{1H}|0\rangle &=T_{1V}|1\rangle=|\varphi_1\rangle\\
T_{2H}|0\rangle &=T_{2V}|1\rangle=|\varphi_2\rangle.
\end{aligned}
\right.
\end{equation}

The initial state for the operations can be represented by
\begin{equation}
|\psi_{in}\rangle = \frac{1}{\sqrt{2}}\left(|\varphi_1\rangle_{1T}|\varphi_2\rangle_{2T}+|\varphi_1\rangle_{1R}|\varphi_2\rangle_{2T}\right).
\end{equation}

The configurable parts in each spatial mode are then applied to the states. In our implementation, the elements placed in modes of $1T$ and $2T$ are fixed to be the projectors in horizontal and vertical polarizations respectively, with the elements in the other two spatial modes reconfigurable. These optical elements turn the state to
\begin{equation}
\left({\rm M}_0|\varphi_1\rangle_{1T}\otimes A|\varphi_2\rangle_{2T}+{\rm M}_1|\varphi_1\rangle_{1R}\otimes B|\varphi_2\rangle_{2T}\right),
\end{equation}
where ${\rm M}_0$ and ${\rm M}_1$ are the projectors to $|0\rangle$ and $|1\rangle$ respectively. The success probability of this step is 1/2. $A$ and $B$ denote the two sets of configurable elements. After this operation, the spatial modes are mixed in the BS parts of the crystals. We post-select one of the ports of each crystal (where the probability amplitude is $\frac{1}{\sqrt{2}}$ for each photon), eliminating the path information, and the state becomes
\begin{equation}
\begin{aligned}
&\left({\rm M}_0|\varphi_1\rangle\otimes A|\varphi_2\rangle+{\rm M}_1|\varphi_1\rangle\otimes B|\varphi_2\rangle\right)\\
=&\left({\rm M}_0\otimes A+{\rm M}_1\otimes B\right)|\varphi_1\rangle|\varphi_2\rangle,
\end{aligned}
\end{equation}
with success probability of 1/4.

Finally, the configurable element $C$ before the detector for the first qubit is applied on this state, and we obtained the final state
\begin{equation}\label{eq:FinalState}
|\psi_{o}\rangle=(C\otimes {\rm I})\left({\rm M}_0\otimes A+{\rm M}_1\otimes B\right)|\varphi_1\rangle|\varphi_2\rangle.
\end{equation}

By configuring $A$, $B$, and $C$, we can realize different operations. It is worth noting that the reflection of the beam on the surfaces of prisms and the PBS/BS cube act as Pauli-Z gates, and can be compensated by a half-wave plate fixed at $0^\circ$.

\subsection{State evolution within the multiply operation}

In multiply operation, the circuit is shown in Fig.~\ref{Fig:ExperimentalProposal}{\color{red}{(c)}}, where the Sagnac interferometers are configured to be a C-NOT gate. Specifically, the parts $A$ and $C$ are configured to identity using half-waveplates fixed at $0^\circ$, and $B$ acts as a Pauli-X gate, via a half-waveplate fixed at $45^\circ$. For simplicity, the states in the following discussion are not necessarily normalized. The initial states are prepared to be
\begin{equation}\label{Eq:InitialState}
\left\{
\begin{aligned}
|\varphi_1\rangle&=h_1|0\rangle+|1\rangle  \\
|\varphi_2\rangle&=h_2|0\rangle+|1\rangle,
\end{aligned}
\right.
\end{equation}
where $h_1$ and $h_2$ are the relative amplitudes of the input states, which are constant numbers or functions of $p$. Theoretically, the relative amplitudes are functions of $p$, while within the experiments, the parameters are usually assigned specific values. Note that the states are not necessarily normalized. We can obtain a final state from equation~(\ref{eq:FinalState}), that is
\begin{equation}\label{Eq:MulFinal}
|\psi_{o1}\rangle=({\rm I}\otimes {\rm I}) \left({\rm M}_0\otimes {\rm I}+{\rm M}_1\otimes {\rm X}\right)|\varphi_1\rangle|\varphi_2\rangle.
\end{equation}
Combine equation~(\ref{Eq:InitialState}) and equation~(\ref{Eq:MulFinal}), the result state is
\begin{equation}
\begin{aligned}
|\psi_{o1}\rangle&=\left({\rm M}_0\otimes {\rm I}+{\rm M}_1\otimes {\rm X}\right)|\varphi_1\varphi_2\rangle\\
&=h_1h_2|00\rangle+h_1|01\rangle+|10\rangle+h_2|11\rangle.\\
\end{aligned}
\end{equation}
We then post-select the second qubit in $|0\rangle$ basis, and the result state collapses to
\begin{equation}\label{Eq:StateMul}
|\psi_{m}\rangle=h_1h_2|0\rangle+|1\rangle.
\end{equation}
Interestingly, if we post-select the second qubit of $|\psi_{o}\rangle$ in $|1\rangle$ basis, we can implement a division operation instead, turning the outcome state to be $|\psi_{d}\rangle=\frac{h_1}{h_2}|0\rangle+|1\rangle$.

\subsection{State evolution within the add operation}

The circuit to implement the add operation is quite complicated, which is to implement a unitary denoted by $B$:
\begin{equation}
B=\left[
\begin{matrix}
1 & 0 & 0 & 0 \\
0 & \frac{1}{\sqrt{2}} & -\frac{1}{\sqrt{2}} & 0 \\
0 & \frac{1}{\sqrt{2}} & \frac{1}{\sqrt{2}} & 0 \\
0 & 0 & 0 & 1
\end{matrix}
\right].
\end{equation}

The circuit requires 3 qubits and 5 control-operations.
\begin{figure}[!htb]
  \centering
  \includegraphics[width=0.65\textwidth]{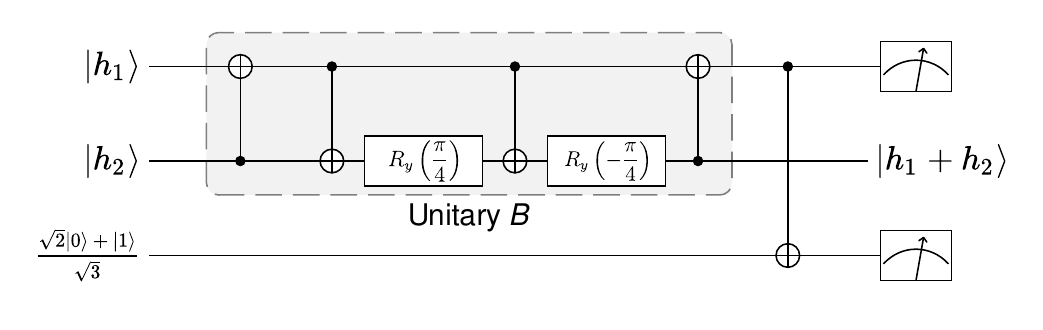}\\
  \caption{\footnotesize{{Circuits for Add Operation By Implementing Unitary $B$}}}\label{Fig:UniB}
\end{figure}

where the unitary $B$ consists of 4 C-NOT gates and 2 single-qubit rotation gates. By applying unitary $B$ on the first 2 qubits, we obtain
\begin{equation}
B|\psi_1\psi_2\rangle=h_1h_2|00\rangle+\frac{h_1-h_2}{\sqrt{2}}|01\rangle+\frac{h_1+h_2}{\sqrt{2}}|10\rangle+|11\rangle.
\end{equation}

The post-selection of the first qubit in $|1\rangle$ basis makes the second qubit collapse into:
\begin{equation}\label{Eq:AddStep2}
|\psi_{o2}\rangle=\frac{h_1+h_2}{\sqrt{2}}|0\rangle+|1\rangle.
\end{equation}

Then multiply $|\psi_{o2}\rangle$ with a constant state $(\sqrt{2}|0\rangle+|1\rangle)$ produces the final state:
\begin{equation}\label{eq:READD}
|\psi_{a}\rangle=(h_1+h_2)|0\rangle+|1\rangle.
\end{equation}

We simplify the circuit, as shown in Fig.~\ref{Fig:ExperimentalProposal}{\color{red}{(d)}} in main text. Practically, the reversion on the first qubit is merged into the initial state preparation, that is, we prepare the initial state to be
\begin{equation}\label{Eq:InitialStateAdd}
\left\{
\begin{aligned}
|\phi_1\rangle&=\frac{1}{h_1}|0\rangle+|1\rangle  \\
|\phi_2\rangle&=h_2|0\rangle+|1\rangle.
\end{aligned}
\right.
\end{equation}

The configurable elements are then reconfigured for the add operation. Specifically, $A$ is configured as identity using a half-waveplate fixed at $0^\circ$, $B$ is the M$_0$X gate by using a half-waveplate fixed at $45^\circ$ and a polarizer that filters photons with horizontal polarization, and $C$ is the Hadamard gate (using a half-waveplate fixed at $22.5^\circ$). Combining with equation~(\ref{eq:FinalState}), the output state before the post-selection is
\begin{equation}
\begin{aligned}
|\psi_{o4}\rangle&={\rm H}\otimes {\rm I}\left({\rm M}_0|\phi_1\rangle\otimes {\rm I}|\phi_2\rangle+{\rm M}_1|\phi_1\rangle\otimes {\rm M}_0{\rm X}|\phi_2\rangle \right)\\
&=(h_1+h_2)|00\rangle+|01\rangle+(h_2-h_1)|10\rangle+|11\rangle.
\end{aligned}
\end{equation}

Then, post-select the first qubit in $|0\rangle$ basis, and we will obtain the final state of $|\psi_a\rangle$ as shown in equation~(\ref{eq:READD}). Similarly, if we post-select the first qubit in $|1\rangle$ basis, we can implement subtract operation instead.

\subsection{Success probability of the operations}

As discussed in the above section, the final state is obtained through several cascades of poet-selections. From the representations of the final output state, we can find that the success probability is different according to different values of $h_1$ and $h_2$.

The success probabilities for the multiply operation and add operation can be calculated by
\begin{equation}\label{Eq:SuccPr}
\left\{
\begin{aligned}
{\Pr}_{m}&=\frac{|h_1|^2|h_2|^2+1}{8(|h_1|^2+1)(|h_2|^2+1)}\\
{\Pr}_{a}&=\frac{|h_1+h_2|^2+1}{16(|h_1|^2+1)(|h_2|^2+1)}.
\end{aligned}
\right.
\end{equation}

It means that for some specific values of $h_1$ and $h_2$, the success probability to obtain the result states become quite low, making it more difficult to evaluate the fidelity of the output state. 
The maximum success probability for multiply operation reaches $\frac{1}{8}$ when $|h_1|=|h_2|=0$ or $|h_1|=|h_2|=\infty$, corresponding to the cases where the initial state is $|HH\rangle$ or $|VV\rangle$. For add operation, the maximum success probability is reached at $\frac{1}{12}$, when $h_1=h_2=\pm\frac{\sqrt{2}}{2}$.

\subsection{Evaluation of C-NOT gate}

We configure the photonic logic to be a C-NOT gate, and then use the method proposed in~\cite{Hofmann2005S} to evaluate the C-NOT gate. The process fidelity of the C-NOT gate can be evaluated through the measurement of two truth tables in complimentary basis and then calculate the classical fidelities of the two truth tables through

\begin{equation}\label{Eq:ClassicalFidelityFormula}
F_P=\frac{1}{N}\sum_{i=1}^N\Pr(f(i)|i),
\end{equation}
where $\Pr(f(i)|i)$ denotes the probability to obtain the theoretical output $f(i)$ when the input $i$ is given. We choose $\{|HH\rangle,|HV\rangle,|VH\rangle,|VV\rangle\}$ and $\{|DD\rangle,|DA\rangle,|AD\rangle,|AA\rangle\}$ as the bases for the fidelity evaluation, thus the classical fidelities can be calculated through
\begin{equation}\label{Eq:ClassicalFidelityTwoTruthTable}
\left\{
\begin{aligned}
F_{HV}&=\frac{\Pr(HH|HH)+\Pr(HV|HV)+\Pr(VH|VV)+\Pr(VV|VH)}{4}\\
F_{DA}&=\frac{\Pr(DD|DD)+\Pr(DA|AA)+\Pr(AD|AD)+\Pr(AA|DA)}{4}.
\end{aligned}
\right.
\end{equation}

The results of the two truth tables of the C-NOT gate in the form of coincidence counts are shown in TABLE~\ref{TAB:TruthTablesCoincidenceCounts}.
\begin{table}[!htb]
  \centering
  \caption{{Truth tables of C-NOT gate}. In both the tables, the first row represents the input basis and the first column represents the output basis. The coincidence counts are accumulated in 50 seconds.}\label{TAB:TruthTablesCoincidenceCounts}
  \begin{tabular}{cp{2cm}c}
      \begin{tabular}{c|rrrr}
      \Xhline{1pt}
         & HH\hspace{0.3cm}   & HV\hspace{0.3cm}    & VH\hspace{0.3cm}    & VV\hspace{0.3cm}   \\
      \hline
      HH & 2061 & 41    & 7     & 0    \\
      HV & 41   & 1826  & 3     & 16   \\
      VH & 14   & 15    & 39    & 1966 \\
      VV & 15   & 7     & 2065  & 26   \\
      \Xhline{1pt}
      \end{tabular}
    &    &
      \begin{tabular}{c|rrrr}
      \Xhline{1pt}
         & DD\hspace{0.3cm}   & DA\hspace{0.3cm}    & AD\hspace{0.3cm}    & AA\hspace{0.3cm}  \\
      \hline
      DD & 1580 & 5     & 105   & 12  \\
      DA & 12   & 100   & 0     & 2060\\
      AD & 95   & 7     & 2132  & 6   \\
      AA & 3    & 1939  & 13    & 117 \\
      \Xhline{1pt}
      \end{tabular}
  \end{tabular}
\end{table}

\begin{figure}[t]
  \centering
  \begin{tabular}{cc}
  \includegraphics[width=0.23\textwidth]{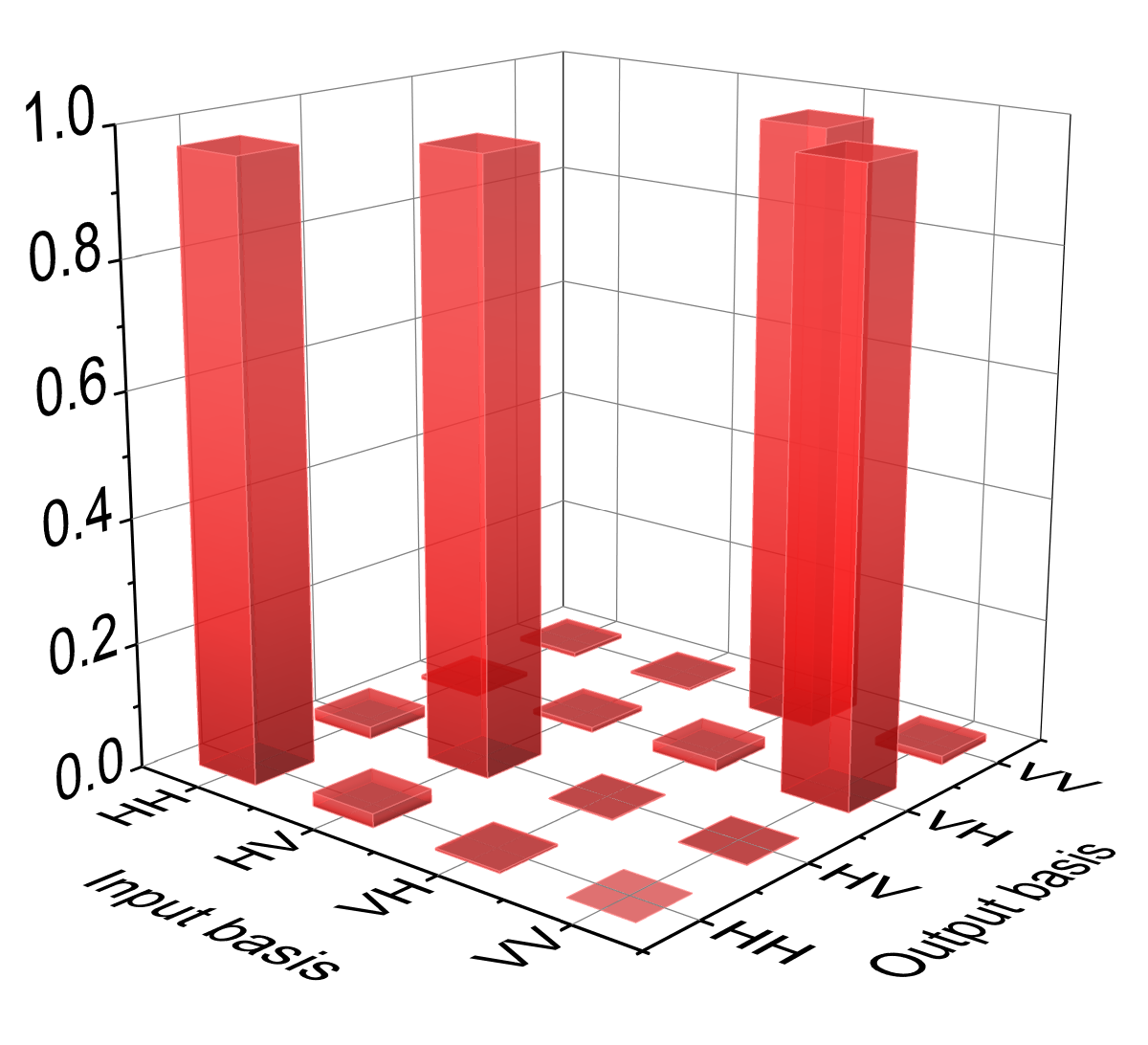}&
  \includegraphics[width=0.23\textwidth]{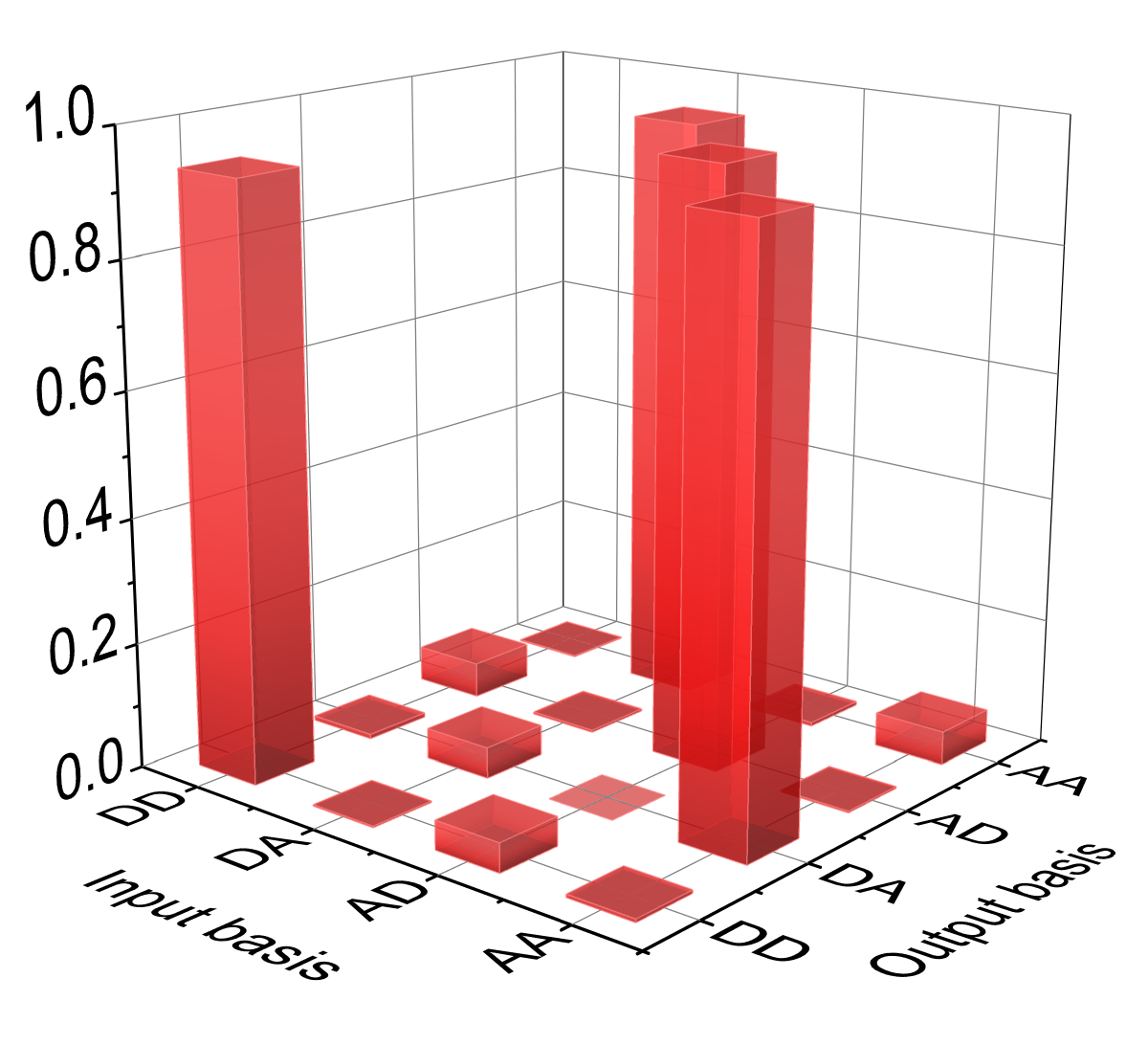}\\
  \end{tabular}
  \caption{\footnotesize{{\bf The results of truth tables of the C-NOT gate.} We choose $\{|HH\rangle, |HV\rangle, |VH\rangle, |VV\rangle\}$ and $\{|DD\rangle, |DA\rangle, |AD\rangle, |AA\rangle\}$ as the bases for fidelity evaluation. Fidelities of the truth tables are $97.24\pm 0.65\%$ ($|H\rangle$/$|V\rangle$) and $94.16\pm0.59\%$ ($|D\rangle$/$|A\rangle$) respectively. The truth tables are obtained through coincidence counts, with each high column being around 2,000.}}\label{Fig:truthtable}
\end{figure}

By converting the coincidence counts into probability, as shown in FIG.~\ref{Fig:truthtable} and TABLE~\ref{TAB:TruthtablesPro}, we can evaluate the classical fidelities of the two truth tables through
%
\begin{table}[!htb]
  \centering
  \caption{{Classical fidelity of the C-NOT gate}. In both the truth tables, the first row represents the input basis and the first column represents the output basis. The data are sued for the bar graphes in Fig.~\ref{Fig:truthtable} in main text}\label{TAB:TruthtablesPro}

  \begin{tabular}{cp{2cm}c}
      \begin{tabular}{c|rrrr}
      \Xhline{1pt}
           & HH\hspace{0.5cm}   & HV\hspace{0.5cm}  & VH\hspace{0.5cm}  & VV\hspace{0.5cm}  \\
      \hline
        HH & 96.72\%            &  2.17\%           &  0.33\%           &  0.00\%           \\
        HV &  1.92\%            & 96.66\%           &  0.14\%           &  0.80\%           \\
        VH &  0.66\%            &  0.79\%           &  1.84\%           & 97.91\%           \\
        VV &  0.70\%            &  0.37\%           & 97.68\%           &  1.29\%           \\
      \Xhline{1pt}
      \end{tabular}
    &    &
      \begin{tabular}{c|rrrr}
      \Xhline{1pt}
           & DD\hspace{0.5cm}   & DA\hspace{0.5cm}  & AD\hspace{0.5cm}  & AA\hspace{0.5cm}  \\
      \hline
        DD & 93.49\%            &  0.24\%           &  4.67\%           &  0.55\%           \\
        DA &  0.71\%            &  4.88\%           &  0.00\%           & 93.85\%           \\
        AD &  5.62\%            &  0.34\%           & 94.76\%           &  0.27\%           \\
        AA &  0.18\%            & 94.54\%           &  0.58\%           &  5.33\%           \\
      \Xhline{1pt}
      \end{tabular}
  \end{tabular}
\end{table}

\begin{equation}\label{Eq:ClassicalFidelityTwoTruthTable}
\left\{
\begin{aligned}
F_{HV}&=\frac{1}{4}\left(P_{HH\rightarrow HH}+P_{HV\rightarrow HV}+P_{VH\rightarrow VV}+P_{VV\rightarrow VH}\right)=97.24\%\\
F_{DA}&=\frac{1}{4}\left(P_{DD\rightarrow DD}+P_{DA\rightarrow AA}+P_{AD\rightarrow AD}+P_{AA\rightarrow DA}\right)=94.16\%.
\end{aligned}
\right.
\end{equation}

The process fidelity of the C-NOT gate can then be bounded by
\begin{equation}\label{Eq:ProcessFidelity}
F_{HV}+F_{DA}-1\leq F_P \leq \min(F_{HV, F_{DA}}),
\end{equation}
and the fidelity over all input states through the average gate fidelity is calculated through
\begin{equation}\label{Eq:AverageFidelity}
\bar{F}=\frac{N\cdot F_P+1}{N+1},
\end{equation}
where $N = 4$ for our 2 qubits system. The fidelities of the C-NOT gate then can be bounded as
\begin{equation}
\left\{
\begin{aligned}
91.40\%\leq &F_P \leq94.16\%\\
93.12\%\leq &\bar{F} \leq95.33\%.\\
\end{aligned}
\right.
\end{equation}

\subsection{Experimental Proposal for $|p\rangle$}

Besides the example of $|f_q(p)\rangle=(2p-1)|0\rangle+|1\rangle$, another important Bernoulli state is $|p\rangle=p|0\rangle+|1\rangle$. Though the corresponding classical coin is classically constructible, this state itself plays an important role in the construction of Bernoulli states~\cite{Jiang2018S}. Note that the states are not necessarily normalized.

%
%
%

The photonic logic can be flexibly configured to generate this Bernoulli state. The photonic logic is first configured as a C-NOT operation, that is, $A$ is configured to be identity, and $B$ is configured to be X. Besides these configurations, several additional optical elements are placed in the control loop: a Hadmard gate is placed after the ${\rm M}_0$ projector in the $1T$ route, and two projectors are placed after the ${\rm M}_1$ projector in the $2T$ route to half the amplitude of the $|1\rangle$ part, resulting that:
\begin{equation}
\begin{aligned}
|\psi_{op}\rangle&=\left({\rm H}{\rm M}_0|\psi_p\rangle_{1T}\right)\otimes \left({\rm H}{\rm M}_0|\psi_p\rangle_{2T}\right)+\left({\rm M}_1{\rm M}_d{\rm M}_1|\psi_p\rangle_{1R}\right)\otimes {\rm X}|\psi_p\rangle_{2T},\\
\end{aligned}
\end{equation}
where M$_d$ is the projector onto state $|+\rangle=\frac{1}{\sqrt{2}}\left(|0\rangle+|1\rangle\right)$. M$_d$ with two ${\rm M}_1$ operators can reduce the amplitude of $|1\rangle$ by a half. After having been mixed at the BS part of the mixed crystal, the state becomes:
\begin{equation}
\begin{aligned}
|\psi_{o}\rangle
&=\frac{1}{2(1-p)}\left(p|0\rangle+|1\rangle\right)\otimes |0\rangle+\frac{p}{2(1-p)}|01\rangle+\left(\frac{p}{2(1-p)}+\frac{1}{2}\sqrt{\frac{p}{1-p}}\right)|11\rangle.
\end{aligned}
\end{equation}
Then if we obtain $|0\rangle$ when measuring the second qubit, the remaining qubit collapses into $|p\rangle$.

\section{Quantum advantage through the example coin}

\label{SEC:QA}
\begin{figure}
  \centering
  \includegraphics[width=0.45\textwidth]{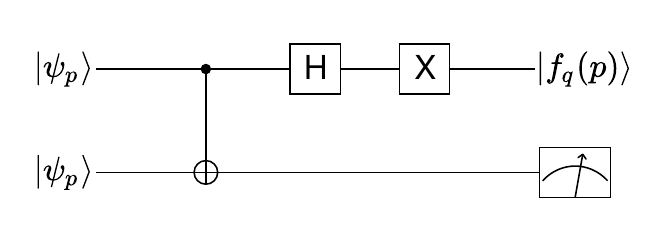}\\
  \caption{\footnotesize{{The quantum circuit for the example quoin.}}}\label{Fig:fpCircuit}
\end{figure}

We use the function $f_c(p)=1-\frac{1}{1+(2p-1)^2}$ as the case to show the advantage of quantum processes. The quantum circuit to generate this state is shown in Fig.~\ref{Fig:fpCircuit}.

Because of $f_c(0.5)=0$, this function is classical infeasible. At the point of $p=0.5$, the quantum advantage can be maximally illustrated.

This circuit can also be realized based on the experimental set up shown in Fig.~\ref{Fig:ExperimentalProposal}(e) in the main text. Specifically, part $A$ and part $B$ are configured so that the sagnac loops act as a C-NOT gate, and part $C$ is consisted of a Hadmard gate and a following Pauli X gate using two waveplates fixed at $22.5^\circ$ and $45^\circ$ respectively. Then, the photonic logic can generate $|f(p)\rangle$ with success probability
\begin{equation}
{\Pr}_c=\frac{(2p-1)^2+1}{16},
\end{equation}
which reaches the minimum value 0.0625 at $p=0.5$. Thus averagely, it requires 16 $|\psi_p\rangle|\psi_p\rangle$ states (32 quoins) to obtain one result coin. However, the resource consumption increases owing to the loss and mode mismatch. To evaluate the loss of transmission, we firstly prepare the state $|\psi_p\rangle|\psi_p\rangle$, and place no optical elements in the sagnac loops, and then accumulate the coincidence counts, which represents the total number of state $|\psi_p\rangle|\psi_p\rangle$ recieved. Then we place the optical elements in, and accumulate the coincidence counts. By comparing the two results of coincidence counts, we found that around 2/5 photons would be lost during the transmission. Thus totally we need about 27 copies of $|\psi_p\rangle|\psi_p\rangle$ states, i.e. $\sim$54 quoins for one $f_c(p)$-coin with $p=0.5$.

Because of the experimental imperfection, the value of the constructed function can not reach 0 when $p=0.5$. This provide the possibility for classical construction. We provide an efficient approach for this construction. The route for constructing $f_c(p)$ is
\begin{equation}
  p \stackrel{s.1}{\Longrightarrow} 2p(1-p) \stackrel{s.2}{\Longrightarrow} 4p(1-p) \stackrel{s.3}{\Longrightarrow} (2p-1)^2 \stackrel{s.4}{\Longrightarrow} \frac{1}{1+(2p-1)^2} \stackrel{s.5}{\Longrightarrow} 1-\frac{1}{1+(2p-1)^2}.
\end{equation}

Step $s.1$ can be finished by tossing the coin twice. In step $s.3$ and step $s.5$, the only thing to do is to change the tail and head defined for the coins. In step $s.4$, it requires to construct a $1/(1+p)$-coin, which can be constructed by the following steps: Toss the $p$-coin twice, if the first toss is tail, then output tail; otherwise if the second toss is tail, output head; otherwise if both tosses are head, repeat this step. The expectation for tossing $p$-coins is

\begin{equation}
  N=2\sum_{i=1}^\infty i(1-p^2)(p^2)^{i-1}=\frac{2}{1-p^2}.
\end{equation}

Then the only difficulty resides on step $s.2$, where it quires to construct a $2p$-coin, which is classical infeasible. We analyze the experimental data, and found that since $p\in[0,1]$, then $2p(1-p)\in[0,0.5]$. let $q = 2p(1-p)$, then we need to construct a $l(q)=Cq$-coin, where $q\in[0,0.5]$, and $C$ is a parameter should be 2 if the experiment is ideal. Practically the parameter $C$ can be obtained by fitting the experimental results, and this function can be classically feasible as long as $C\leq 2$. According to the experimental results, we infer the corresponding results obtained for $Cq$-coin as shown in Tab.~\ref{Tab:CorrespondingDataForC}.

\begin{table}[htb]
\centering
\caption{\footnotesize{\bf Corresponding data for the linear function $l(q)=Cq$ to construct.}}
\label{Tab:CorrespondingDataForC}
\begin{tabular}{cccc}
\hline
  \hspace{1cm}$p$\hspace{1cm}   & $q$($=2p(1-p)$)  &\hspace{1cm}$f_{ct}(p)$\hspace{1cm} &   $Cq$($=C\cdot2p(1-p)$) \\
\hline
    0.0 & 0.00 & 0.483 & 0.066 \\
    0.1 & 0.18 & 0.394 & 0.350 \\
    0.2 & 0.32 & 0.285 & 0.601 \\
    0.3 & 0.42 & 0.161 & 0.808 \\
    0.4 & 0.48 & 0.083 & 0.909 \\
    0.5 & 0.50 & 0.042 & 0.956 \\
    0.6 & 0.48 & 0.089 & 0.902 \\
    0.7 & 0.42 & 0.196 & 0.756 \\
    0.8 & 0.32 & 0.310 & 0.551 \\
    0.9 & 0.18 & 0.431 & 0.243 \\
    1.0 & 0.00 & 0.496 & 0.016 \\
\hline
\end{tabular}
\end{table}

Since it appears the duplicated value of $q$, we take the average value of $Cq$, and by fitting the data we have $C\sim1.868$ and it satisfies $2Cp(1-p)<1-0.066$ for $2p(1-p)\leq0.5$. Now we accumulate the coin-consumption in each step. Step $s.1$ requires 2 coins for a 2p(1-p) coin; Step  $s.2$  requires requires $9.5C/0.066=268.879$ coins by following the evaluation in ref.~\cite{Huber2016S}; Step  $s.3$ and step  $s.5$  requires no coins; Step  $s.4$  requires $2\times1/(1-0.044^2)=2.009$. In total, the consumption of classical coins is $2\times268.879\times2.009\approx1.080\times 10^3$. The quantum advantage is clearly shown here.

\section{Detailed Data}

We identified the quality of the output states by measuring its fidelity. Instead of doing a complete state tomography, we directly measure its fidelity because the theoretical output is already known, and we don't need other information contained in its density matrix. By measuring the counts of photons in the bases that parallel and orthogonal to the polarization of the theoretical state, we can evaluate the fidelity quickly. The data of the experiments are shown in TABLE~\ref{TAB:DataMul}$\sim$\ref{TAB:DataExCoin}.

\begin{table*}[!htb]
  \centering
  \caption{\footnotesize{{{Results of the multiply operation.}} The initial state is prepared to be $(h_1|0\rangle+|1\rangle)\otimes (h_2|0\rangle+|1\rangle)$. The theoretical output state is $(h_1h_2|0\rangle+|1\rangle)$. Note that the normalizing coefficients are not shown for simplicity. The labels correspond to the bars in Fig.~\ref{Fig:FidelityMA} in main text. Terms denoted by ``R$x$M'' or ``C$x$M'' are random generated real or complex numbers. The coincidence counts are accumulated in 10 seconds, except for the ``C$x$M'' cases where the time accumulated is about 50 seconds  and the ``L$x$M'' cases where the time accumulated is about 25 seconds. $CC_\parallel$ is the coincidence counts obtained by setting the measurement basis parallel to the output state, and $CC_\perp$ is the coincidence counts obtained by setting the measurement basis perpendicular to the output state.}}\label{TAB:DataMul}
  \begin{tabular}{crrrrrrrr}
  \Xhline{1pt}
  Label & $\hspace{0.2cm}h_1\hspace{0.8cm}$ & $\hspace{0.2cm}h_2\hspace{0.8cm}$ & $\hspace{0.2cm}h_1\cdot h_2\hspace{0.5cm}$ & \hspace{0.3cm}$CC_\parallel$ & \hspace{0.3cm}$CC_\perp$ & \hspace{0.3cm}Fidelity\hspace{0.1cm} & \hspace{0.3cm}Std.dev\hspace{0.1cm} \\
  \hline
    D1M & 		  $1.000$\hspace{0.6cm}	& 		 $ 1.000$\hspace{0.6cm}	& 		 $ 1.000$\hspace{0.6cm}	&  514  & 20 	&  96.255\% & 0.040\%\\
    D2M & 		  $1.000$\hspace{0.6cm} & 		 $-1.000$\hspace{0.6cm}	& 		 $-1.000$\hspace{0.6cm}	&  455  & 26 	&  94.595\% & 0.039\% \\
    D3M & 		 $-1.000$\hspace{0.6cm}	& 		 $ 1.000$\hspace{0.6cm}	& 		 $-1.000$\hspace{0.6cm}	&  636  & 23 	&  96.510\% & 0.031\% \\
    D4M & 		 $-1.000$\hspace{0.6cm}	& 		 $-1.000$\hspace{0.6cm}	& 		 $ 1.000$\hspace{0.6cm}	&  562  & 23 	&  96.068\% & 0.034\% \\
    H1M & 		  $0.000$\hspace{0.6cm}	& 		  $0.000$\hspace{0.6cm}	& 		  $0.000$\hspace{0.6cm}	&  893  &  1	&  99.888\% & 0.112\% \\
    H2M & 		  $0.000$\hspace{0.6cm}	& 		  $1.000$\hspace{0.6cm}	& 		  $0.000$\hspace{0.6cm}	&  513  &  4	&  99.226\% & 0.096\% \\
    H3M & 		  $0.000$\hspace{0.6cm}	& 		  $5.000$\hspace{0.6cm}	& 		  $0.000$\hspace{0.6cm}	&   38  &  7	&  84.444\% & 0.711\% \\
    H4M & 		  $0.000$\hspace{0.6cm}	& 		  $10.00$\hspace{0.6cm}	& 		  $0.000$\hspace{0.6cm}	&   36  &  8	&  81.818\% & 0.661\% \\
    L1M & 			 $ i$\hspace{0.6cm}	& 			 $ i$\hspace{0.6cm}	& 		 $-1.000$\hspace{0.6cm}	& 1074  & 49 	&  95.637\% & 0.012\% \\
    L2M & 			 $ i$\hspace{0.6cm}	& 			 $-i$\hspace{0.6cm}	& 		 $ 1.000$\hspace{0.6cm}	&  851  & 66 	&  92.803\% & 0.012\% \\
    L3M & 			 $-i$\hspace{0.6cm}	& 			 $ i$\hspace{0.6cm}	& 		 $ 1.000$\hspace{0.6cm}	&  878  & 51 	&  94.510\% & 0.014\% \\
    L4M & 			 $-i$\hspace{0.6cm}	& 			 $-i$\hspace{0.6cm}	& 		 $-1.000$\hspace{0.6cm}	&  811  & 55 	&  93.649\% & 0.015\% \\
    R1M & 		  $0.663$\hspace{0.6cm}	& 		  $0.682$\hspace{0.6cm}	& 		  $0.452$\hspace{0.6cm}	&  579  & 18	&  96.985\% & 0.038\% \\
    R2M & 		  $0.700$\hspace{0.6cm}	& 		  $0.900$\hspace{0.6cm}	& 		  $0.630$\hspace{0.6cm}	&  490  & 20	&  96.078\% & 0.042\% \\
    R3M & 		  $0.080$\hspace{0.6cm}	& 		  $0.830$\hspace{0.6cm}	& 		  $0.066$\hspace{0.6cm}	&  654  &  0	& 100.000\% & 0.000\% \\
    R4M & 		  $0.217$\hspace{0.6cm}	& 		  $0.467$\hspace{0.6cm}	& 		  $0.101$\hspace{0.6cm}	&  654  &  0	& 100.000\% & 0.000\% \\
    R5M & 		  $0.024$\hspace{0.6cm}	& 		  $0.719$\hspace{0.6cm}	& 		  $0.017$\hspace{0.6cm}	&  535  &  0	& 100.000\% & 0.000\% \\
    C1M & $-0.080-0.093i$	& $-0.553-0.821i$	& $-0.032+0.117i$	&  938  & 14	&  98.529\% & 0.028\% \\
    C2M & $ 1.354-1.693i$	& $ 2.455-1.979i$	& $-0.025-6.837i$	& 1848  & 32	&  98.298\% & 0.009\% \\
    C3M & $-0.385-0.934i$	& $-0.050-0.155i$	& $-0.126+0.106i$	&  818  & 32	&  96.235\% & 0.020\% \\
    C4M & $-0.172-0.784i$	& $-0.019-0.353i$	& $-0.273+0.076i$	& 1018  & 32	&  96.952\% & 0.016\% \\
    C5M & $ 0.876+0.182i$	& $-0.184-0.893i$	& $-0.001-0.816i$	& 1225  & 76	&  94.158\% & 0.008\% \\
    C6M & $ 1.611-1.658i$	& $ 1.119-1.065i$	& $ 0.037-3.572i$	&  858  & 34	&  96.188\% & 0.018\% \\
    C7M & $ 1.229+0.240i$	& $-0.064-0.255i$	& $-0.017-0.329i$	&  889  & 46	&  95.080\% & 0.015\% \\
   \Xhline{1pt}
  \end{tabular}
\end{table*}

\begin{table*}[!htb]
  \centering
  \caption{\footnotesize{{{Results of the add operation.}} The initial state is prepared to be $(h_1|0\rangle+|1\rangle)\otimes (h_2|0\rangle+|1\rangle)$. The theoretical output state is $((h_1+h_2)|0\rangle+|1\rangle)$. Note that the normalizing coefficients are not shown for simplicity. ``$\infty$'' in the table indicates the state of horizontal polarization. Terms denoted by ``R$x$A'' or ``C$x$A'' are random generated real or complex numbers for test. The coincidence counts are accumulated in about 10 seconds except for the ``C$x$A'' cases and ``L$x$A'' cases where the time accumulated is about 50 seconds. $CC_\parallel$ is the coincidence counts obtained by setting the measurement basis parallel to the output state, and $CC_\perp$ is the coincidence counts obtained by setting the measurement basis perpendicular to the output state.}}\label{TAB:DataAdd}
  \begin{tabular}{crrrrrrr}
  \Xhline{1pt}
  Label & $\hspace{0.2cm}h_1\hspace{0.8cm}$ & $\hspace{0.2cm}h_2\hspace{0.8cm}$ & $\hspace{0.2cm}h_1+ h_2\hspace{0.5cm}$ & \hspace{0.3cm}$CC_\parallel$ & \hspace{0.3cm}$CC_\perp$ & \hspace{0.3cm}Fidelity\hspace{0.1cm} & \hspace{0.3cm}Std.dev\hspace{0.1cm}\\
  \hline
    D1A & 		 $ 1.000$\hspace{0.6cm} & 		 $ 1.000$\hspace{0.6cm}	&  		  $ 2.000$\hspace{0.6cm}	& 291 &  6 &  97.980\% & 0.135\% \\
    D2A & 		 $-1.000$\hspace{0.6cm} & 		 $-1.000$\hspace{0.6cm}	&  		  $-2.000$\hspace{0.6cm}	& 453 & 15 &  96.795\% & 0.053\%  \\
    D3A & 		 $-1.000$\hspace{0.6cm} & 		 $ 1.000$\hspace{0.6cm}	&  		  $ 0.000$\hspace{0.6cm}	& 127 & 38 &  76.970\% & 0.077\%  \\
    H1A & 		 $ 0.000$\hspace{0.6cm} & 		 $ 0.000$\hspace{0.6cm}	&  		  $ 0.000$\hspace{0.6cm}	& 401 & 12 &  97.094\% & 0.068\%  \\
    H2A & 		 $\infty$\hspace{0.6cm}	& 		 $ 0.010$\hspace{0.6cm}	&  		  $\infty$\hspace{0.6cm}	& 160 &  6 &  96.386\% & 0.237\%  \\
    H3A & 		 $\infty$\hspace{0.6cm}	& 		 $ 0.100$\hspace{0.6cm}	&  		  $\infty$\hspace{0.6cm}	&  46 &  5 &  90.196\% & 0.791\%  \\
    H4A & 		  $0.010$\hspace{0.6cm}	& 		 $\infty$\hspace{0.6cm}	&  		  $\infty$\hspace{0.6cm}	& 315 &  0 & 100.000\% & 0.000\%  \\
    H5A & 		  $0.500$\hspace{0.6cm}	& 		  $0.500$\hspace{0.6cm}	&  		   $1.000$\hspace{0.6cm}	& 480 & 14 &  97.166\% & 0.053\%  \\
    L1A & 		 	 $ i$\hspace{0.6cm}	& 			 $ i$\hspace{0.6cm}	&  		     $ 2i$\hspace{0.6cm}	& 741 & 15 &  98.016\% & 0.033\%  \\
    L2A & 		 	 $-i$\hspace{0.6cm}	& 			 $-i$\hspace{0.6cm}	&  		     $-2i$\hspace{0.6cm}	& 707 & 23 &  96.849\% & 0.028\%  \\
    R1A & 		 $-5.347$\hspace{0.6cm}	& 		 $-3.168$\hspace{0.6cm}	&  		  $-8.515$\hspace{0.6cm}	& 106 &  3 &  97.248\% & 0.515\%  \\
    R2A & 		 $-8.166$\hspace{0.6cm}	& 		 $-0.945$\hspace{0.6cm}	&  		  $-9.111$\hspace{0.6cm}	& 238 &  5 &  97.942\% & 0.180\%  \\
    R3A & 		 $-2.140$\hspace{0.6cm}	& 		 $-1.881$\hspace{0.6cm}	&  		  $-4.021$\hspace{0.6cm}	& 313 &  2 &  99.365\% & 0.223\%  \\
    R4A & 		 $-1.418$\hspace{0.6cm}	& 		 $-6.335$\hspace{0.6cm}	&  		  $-7.753$\hspace{0.6cm}	& 187 &  0 & 100.000\% & 0.000\%  \\
    R5A & 		 $-7.123$\hspace{0.6cm}	& 		 $ 0.038$\hspace{0.6cm}	&  		  $-7.085$\hspace{0.6cm}	& 241 &  6 &  97.571\% & 0.161\%  \\
    R6A & 		 $ 0.256$\hspace{0.6cm}	& 		 $-1.125$\hspace{0.6cm}	&  		  $-0.869$\hspace{0.6cm}	& 275 &  8 &  97.173\% & 0.121\%  \\
    C1A & $-0.400+2.288i$	& $ 0.336+0.948i$	&  $-0.065+3.237i$	& 714 & 17 &  97.674\% & 0.032\%  \\
    C2A & $-0.693+2.360i$	& $ 0.595+1.105i$	&  $-0.096+3.465i$	& 466 & 10 &  97.899\% & 0.065\%  \\
    C3A & $-0.148+1.188i$	& $ 4.329-1.157i$	&  $ 4.181+0.016i$	& 283 &  1 &  99.648\% & 0.351\%  \\
    C4A & $ 0.853+1.024i$	& $ 1.945-0.880i$	&  $ 2.801+0.143i$	& 312 & 12 &  96.296\% & 0.086\%  \\
    C5A & $ 0.184+0.035i$	& $ 1.294+1.030i$	&  $ 6.543+0.041i$	& 502 & 35 &  93.482\% & 0.029\%  \\
    C6A & $-0.338+0.836i$	& $ 0.309+0.981i$	&  $-0.028+1.819i$	& 831 & 20 &  97.650\% & 0.026\%  \\
    C7A & $ 0.794-0.024i$	& $ 0.904-0.080i$	&  $ 1.699-0.106i$	& 972 & 20 &  97.984\% & 0.022\%  \\
    C8A & $ 0.136+0.090i$	& $-0.119-0.911i$	&  $ 0.018-0.823i$	& 578 & 17 &  97.143\% & 0.040\%  \\
    C9A & $-0.928+0.905i$	& $ 0.651-0.618i$	&  $ 0.100-1.156i$	& 285 & 17 &  94.371\% & 0.076\%  \\
  \Xhline{1pt}
  \end{tabular}
\end{table*}

\begin{table*}[!htb]
  \centering
  \caption{\footnotesize{{{Detailed data for the produced function $f_c(p)=1-\frac{1}{1+(2p-1)^2}$.}} The initial state is prepared in $|\psi_p\rangle|\psi_p\rangle=(\sqrt{\frac{p}{1-p}}|0\rangle+|1\rangle)\otimes (\sqrt{\frac{p}{1-p}}|0\rangle+|1\rangle)$. We firstly construct a quoin represented by $|f_q(p)\rangle=(2p-1)|0\rangle+|1\rangle$ (not necessarily normalized), and measure its fidelity. By measuring this quoin in $\sigma_z$ basis, we obtain the classical coin that presents head in probability of $f_c(p)$. The coincidence counts are all accumulated in about 1 minute.
  }}\label{TAB:DataExCoin}
  \begin{tabular}{crrcrrccc}
  \Xhline{1pt}
  $p$ & $CC_\parallel$ & $CC_\perp$ & Fidelity & $CC_H$ & $CC_V$ & $\text{Pr}_{\text{Theor.}}$ & $\text{Pr}_{\text{Exp.}}$ & $\text{Std.deviation }$ \\
  \hline
  0.0	& 1086	&    3 &  99.725\% & 1589 & 1704 & 0.500 & 0.483 & $1.025\times 10^{-5}$ \\
  0.1	&  809	&    7 &  99.142\% &  998 & 1538 & 0.390 & 0.394 & $1.408\times 10^{-5}$ \\
  0.2	&  663	&   14 &  97.932\% &  594 & 1485 & 0.265 & 0.285 & $2.036\times 10^{-5}$ \\
  0.3	&  515	&   15 &  97.170\% &  250 & 1294 & 0.138 & 0.161 & $4.111\times 10^{-5}$ \\
  0.4	&  486	&   17 &  96.620\% &  118 & 1299 & 0.038 & 0.083 & $6.499\times 10^{-5}$ \\
  0.5	&  388	&   14 &  96.517\% &   51 & 1151 & 0.000 & 0.042 & $1.165\times 10^{-4}$ \\
  0.6	&  390	&   12 &  97.015\% &  112 & 1141 & 0.038 & 0.089 & $7.545\times 10^{-5}$ \\
  0.7	&  423	&    7 &  98.372\% &  237 &  978 & 0.138 & 0.196 & $5.384\times 10^{-5}$ \\
  0.8	&  505	&    9 &  98.249\% &  472 & 1049 & 0.265 & 0.310 & $3.161\times 10^{-5}$ \\
  0.9	&  595	&    1 &  99.832\% &  776 & 1025 & 0.390 & 0.431 & $2.387\times 10^{-5}$ \\
  1.0	&  751	&    1 &  99.987\% & 1109 & 1130 & 0.500 & 0.496 & $1.871\times 10^{-5}$ \\
  \Xhline{1pt}
  \end{tabular}
\end{table*}

\end{document}